\documentstyle[12pt]{article}

\begin{document}
\begin{titlepage}
\begin{center}
July 21, 1998     \hfill    LBNL-40722 \\

\vskip .2in

{\large \bf Quantum Ontology and Mind-Matter Synthesis}
\footnote{This work was supported by the Director, Office of Energy 
Research, Office of High Energy and Nuclear Physics, Division of High 
Energy Physics of the U.S. Department of Energy under Contract 
DE-AC03-76SF00098.}
\vskip .2in
Henry P. Stapp\\
{\em Lawrence Berkeley National Laboratory\\
      University of California\\
    Berkeley, California 94720}
\end{center}

\begin{abstract}

The  Solvay  conference of 1927  marked  the birth of  quantum  theory. This
theory constitutes a radical break  with prior tradition in physics, because
it avers, if taken seriously, that nature is built not out of matter but out
of knowings.  However, the  founders of the  theory  stipulated, cautiously,
that  the  theory  was not  to be  taken   seriously,  in  this  sense, as a
description of  nature herself, but  was to be construed  as merely a way of
computing  expectations about  future  knowings on the  basis of information
provided by past knowings. There have been many efforts over the intervening
seventy years to rid physics of this  contamination of matter by mind. But I
use  the  reports  at  this   Symposium to   support the   claim  that these
decontamination     efforts  have   failed,  and  that,   because of  recent
developments  pertaining  to  causality, the time  has come to  take quantum
theory seriously: to  take it as the basis for a  conception of the universe
built on knowings, and other things of the same kind. Quantum theory ensures 
that this  conception will yield all the empirical regularities that  had 
formerly been thought to arise from the   properties of   matter,  together  
with  all of  those  more  recently discovered   regularities  that  cannot 
be understood in that mechanical way. Thus I propose to break away from the 
cautious stance of the  founders of quantum theory, and build  a  theory of 
reality by taking seriously what the incredible accuracy of the predictions 
of the formalism seems to proclaim, namely that nature is best understood as  
being built around knowings that enjoy the mathematical properties ascribed 
to them by  quantum theory. I explain why this idea had formerly been 
incorrectly regarded as untenable, due to a failure  to distinguish signals 
from influences:  relativistic quantum field theory ensures both that signals 
cannot travel faster than light,  but that influences, broadly conceived, 
cannot be imagined to enjoy that  property. Failure to  recognize this fact 
had made a realistic interpretation of  quantum  theory seem  impossible. I 
then explain how our conscious knowings can play a causally efficacious and 
binding role in brain dynamics without violating the statistical rules of 
quantum theory, and  describe how these features provide a  foundation for 
understanding how consciousness could have evolved by natural selection from 
primitive beginnings.

\end{abstract}
\begin{center}
Invited Paper: The X-th Max Born Symposium ``Quantum Future''.
\end{center}
\end{titlepage}

\newpage
\renewcommand{\thepage}{\arabic{page}}
\setcounter{page}{1}
\noindent {\bf 1. Introduction.}

The  modern era  was  created  probably  as much  by  Descartes'  conceptual
separation  of  mind from  matter as  by any  other  event. This  move freed
science  from  the religious  dogmas and  constraints of  earlier times, and
allowed  scientists to delve into the important mathematical regularities of
the  observed physical world.  Descartes himself allowed interaction between
mind  and  matter to  occur within  the confines  of a human  brain, but the
deterministic character  of the physical world  specified later by Newtonian
mechanics   seemed to  rule out   completely,  even within  our  brains, any
interference  of  mind with the  workings  of matter.  Thus the  notion of a
completely   mechanical  universe,  controlled by  universal  physical laws,
became the new dogma of science.

It can readily be imagined that within the milieu dominated by such thinking
there would be  stout  opposition to the  radical claims of  the founders of
quantum  theory that our  conscious  human  knowings should be  taken as the
basis of  our  fundamental  theory of  nature.  Yet the   opposition to this
profound  shift  in  scientific  thinking  was  less  fierce than  one might
suppose. For, in the end, no one could dispute that science rests on what we
can know, and  quantum theory was  formulated in  practical human terms that
rested squarely on  that fact. Hence  the momentous  philosophical shift was
achieved by some subtle linguistic  reformulations that were inculcated into
the minds  of the  students and   practitioners of  quantum  theory. The new
thought patterns, and the  calculations they engendered, worked beautifully,
insofar as one kept to the specified practical issues, and refrained, as one
was  instructed to  do, from  asking  certain  ``meaningless''  metaphysical
questions.

Of  course, there  are a few  physicists  who are  dissatisfied  with purely
practical  success, and  want to  understand what  the  practical success of
these  computational rules is telling  us about  ourselves and the nature of
the world  in  which we live. Efforts  to achieve such  an understanding are
proliferating, and the present work  is of that genre. Historically, efforts
to achieve   increasingly coherent  and comprehensive  understandings of the
clues we extract  from Nature have occasionally led to scientific progress.

The outline of the present work is  as follows. In section 2, I document the
claim made  above that  the orthodox  Copenhagen   interpretation of quantum
theory is based  squarely and  explicitly on human  knowings. The aim of the
paper is  to imbed  this  orthodox  pragmatic   epistemological  theory in a
rationally coherent naturalistic  ontology in a minimalistic way that causes
no  disruption of  anything that  orthodox  quantum theory  says, but merely
supplies  a  natural   ontological   underpinning.  In the  special  case of
processes occurring in human body/brains this ontological structure involves
human conscious knowings that enter into the brain dynamics in a manner that
accounts   for  the  way  that  these   knowings  enter  into  the  orthodox
interpretation of quantum theory.

In section 3 I discuss another  interpretation, which is probably the common
contemporary  interpretation of the Copenhagen  interpretation. It is coarse
in  that it  is  imprecise on  essential  theoretical points.  Because it is
common and coarse I call it the Vulgar Copenhagen Interpretation.

In section 4 the  unusual causal  structure of quantum  theory is discussed,
and is used to justify,  in the context of trying  to understand the role of
mind  in  nature:  1) the   rejection of  the   classical  ontology,  2) the
reasonableness of  attempting to ontologicalize  the orthodox interpretation
of  quantum  theory,  and 3)  the   expectation that  our  knowings  involve
non-local aspects. 

Section 5 is entitled ``All roads lead to Solvay 1927''.  The 1927
Solvay conference, seventy years ago, marked the birth of the orthodox
Copenhagen interpretation of quantum theory. In this section I review
this Symposium from a certain point of view, namely the viewpoint that
many of the highlights of the Symposium confirm the basic message of
the orthodox interpretation, namely that the only reasonable way to
make rational sense out of the empirical data is to regard nature as
being built out of knowings. I argue that the experience of the last
seventy years suggests the reasonableness of taking this
interpretation seriously: more seriously than the founders of quantum
theory took it. Basically, they said, cautiously, that the
mathematical formalism is a useful tool for forming expectations about
our future knowings on the basis of our past ones. That claim has been
now been abundantly confirmed, also in fields far beyond the narrow
confines of atomic physics. But the founders scrupulously avoided any
suggestion that this mathematical formalism corresponded to
reality. They either discouraged us from asking questions about what
is really happening, or, if pressed, looked for reality not in their
own knowledge-based formalism, but in terms of more conventional
physical terms.  This reluctance to take their own formalism seriously
was, I think, the result partly of an inertial carry-over from
classical physics, which shunned and excluded any serious
consideration of mind in physics, and partly of a carry-over of an
idea from the special theory of relativity.  This is the idea that no
influence or signal could propagate faster than light.  However, in
quantum theory there is a sharp distinction between signal and
influence, because it can be proved both that no signal can be
transmitted faster than light, and that this property cannot be
imagined to hold for influences.  The distinction between signal and
influence has to do with the difference between the causal structure
of the deterministic evolution of the {\it statistical predictions of
the theory} and the causal structure of something that has no analog
in classical mechanics, namely the {\it selection process} that acts
within the deterministic structure that is the analog of the classical
deterministic structure, but that is not fully determined by that
structure.

In cosmological solutions in general relativity there is usually a preferred
set of  advancing spacelike  surfaces that  provide a  natural definition of
instantaneousness. Also, there is the empirical cosmological preferred frame
defined  by the  background  black-body  radiation.  So the idea  of special
relativity that  there is no  preferred frame for the  universe, although it
may  indeed hold for  the  formulation of  the  general  local-deterministic
laws, is not as  compelling now as it  was in 1905, or  even 1927: that idea
could very well  break down in our  particular universe  at the level of the
selection of particular individual results (knowings). Indeed, I believe  it
{\it must} break down at that level. (Stapp, 1997) 

So I propose to  take seriously the  message of Solvay  1927, that nature be
understood as  built out of knowings.  But we must then  learn how better to
understand   knowings, within  the  mathematical  framework  provided by the
quantum formalism.

In  section 6 I  distinguish  the two  different  components of  the quantum
mechanical evolutionary process, the unitary/local part and the  nonunitary/
nonlocal part, and  note that our  conscious knowings,  as they occur in the
quantum  description,  enter only  into the  latter part. But  that  part is
eliminated   when one  takes the   classical   approximation to  the quantum
dynamics.  Thus  from  the  perspective  of  quantum  mechanics it  would be
irrational   to  try to  find  consciousness  in a  classical  conception of
nature, because that conception corresponds to an approximation to the basic
dynamics  from which  the  process  associated with   consciousness has been
eradicated.

I note  there also that  the  ontologicalization  of the  quantum mechanical
description  dissolves,  or at least  radically  transforms  the mind-matter
dualism. The  reason is this: in the  classical theory  one specifies at the
outset that the mathematical quantities of the theory represent the physical
configuration of matter, and hence  one needs to explain later how something
so seemingly different from matter as  our conscious knowings fit in. But in
the  quantum  case one   specifies from  the  outset that  the  mathematical
quantities of  the theory  describe  properties of  knowings, so there is no
duality that needs explaining: no  reality resembling the substantive matter
of classical physics ever enters at all. One has, instead, a sequence of
events that are associated from the outset with experiences, and that 
evolve within a mathematically specified framework. 

Section 7 lays out more explicitly the two kinds of processes by showing how
they can be  considered to be  evolutions in  two different  time variables,
called process time and mathematical time.

Section 8 goes into the question of the ontological nature of the  ``quantum
stuff'' of the universe.

In the   sections 9 and  10 I  describe  the  proposed  ontology.  It brings
conscious  knowings  efficaciously  into quantum  brain  dynamics. The basic
point is that in a theory with objectively real quantum jumps, some of which
are   identifiable  with  the  quantum  jumps  that  occur in  the  orthodox
epistemological interpretation, one  needs three things that lie beyond what
orthodox quantum theory provides:

\noindent 1.  A process that defines the conditions under which these jumps 
occur, and the possibilities for what that jump might be. 

\noindent 2.  A process that selects which one of the possibilities actually
occurs.

\noindent 3. A process that brings the entire universe into concordance with
the selected outcome.

Nothing in the normal  quantum description of  nature in terms of vectors in
Hilbert space accomplishes either 1 or 2. And 3 is simply put in by hand. So
there  is a  huge  logical  gap in  the  orthodox  quantum   description, if
considered from an  ontological point  of view. {\it  Some extra process, or
set of processes, not described in the orthodox physical theory, is needed.}

I take a  minimalistic and  naturalistic  stance, admitting  only the  least
needed to  account  for the  structure of  the  orthodox quantum  mechanical
rules.

In appendix A I show why the quantum character of certain synaptic processes
make it virtually certain that the quantum collapse process will exercise 
dominant control over the course of a conscious mind/brain processes. 	

\vskip .1in
\noindent {\bf 2. The subjective character of the orthodox interpretation of 
quantum mechanics.}

In  the   introduction to  his  book  ``Quantum   theory and   reality'' the
philosopher   of science  Mario Bunge  (1967)  said: ``The  physicist of the
latest  generation  is   operationalist all  right, but  usually he does not
know, and refuses to   believe, that the original  Copenhagen interpretation
---  which he  thinks  he  supports  --- was  squarely   subjectivist, i.e.,
nonphysical.''

Let there be no doubt about this.

Heisenberg (1958a): ``The conception  of objective reality of the elementary
particles has thus evaporated not into the cloud of some obscure new reality
concept but into the transparent clarity of a mathematics that represents no
longer  the   behavior  of  particles  but   rather our   knowledge  of this
behaviour.''

Heisenberg (1958b):  ``...the act of registration  of the result in the mind
of  the observer. The  discontinuous change in  the probablitity function...
takes  place with the  act of registration,  because it is the discontinuous
change in  our knowledge  in the instant of  registration that has its image
in the discontinuous change of the probability function.'' 

Heisenberg (1958b:) ``When old adage  `Natura non facit saltus' is used as a
basis of a  criticism of  quantum  theory, we can  reply that  certainly our
knowledge can change  suddenly, and that this  fact justifies the use of the
term `quantum jump'. ''

Wigner (1961): ``the laws of quantum mechanics cannot be formulated...without
recourse to the concept of consciousness.''

Bohr (1934): ``In our  description of nature the  purpose is not to disclose
the real  essence of  phenomena but  only to  track down as  far as possible
relations between the multifold aspects of our experience.''

In  his book  ``The   creation of   quantum  mechanics  and  the  Bohr-Pauli
dialogue'' (Hendry, 1984) the historian John Hendry gives a detailed account
of the  fierce struggles by such  eminent thinkers as Hilbert, Jordan, Weyl,
von Neumann, Born,  Einstein,  Sommerfeld, Pauli,  Heisenberg, Schroedinger,
Dirac, Bohr and others, to come up  with a rational way of comprehending the
data from atomic experiments. Each  man had his own bias and intuitions, but
in spite   of  intense effort  no rational   comprehension was  forthcoming.
Finally, at the 1927  Solvay conference a group  including Bohr, Heisenberg,
Pauli, Dirac, and  Born come into  concordance on a solution that came to be
called  ``The    Copenhagen    Interpretation''.  Hendry  says:  ``Dirac, in
discussion, insisted on  the restriction of the  theory's application to our
knowledge  of a system,  and on its  lack of  ontological  content.'' Hendry
summarized  the  concordance by   saying:  ``On this   interpretation it was
agreed  that,  as  Dirac  explained,  the   wave  function   represented our
knowledge  of the  system, and the  reduced wave   packets our  more precise
knowledge after measurement.''

Certainly this profound shift in  physicists' conception of the basic nature
of their endeavour, and the meanings  of their formulas, was not a frivolous
move: it was a last resort. The very idea that in order to comprehend atomic
phenomena one must abandon physical  ontology, and construe the mathematical
formulas to be directly about the  knowledge of human observers, rather than
about the external real events themselves, is so seemingly preposterous that
no group of eminent and renowned  scientists would ever embrace it except as
an extreme last measure. Consequently, it would be frivolous of us simply to
ignore a  conclusion so hard won and  profound, and of  such apparent direct
bearing on our  effort to understand  the connection of  our knowings to our
bodies.

Einstein never accepted the  Copenhagen interpretation. He said: ``What does
not satisfy  me, from the  standpoint of  principle, is its  attitude toward
what seems to  me to be the  programmatic  aim of all  physics: the complete
description  of any  (individual)  real situation  (as it  supposedly exists
irrespective  of any act  of  observation of   substantiation).'' (Einstein,
1951,   p.667) and  ``What I dislike  in this  kind of  argumentation is the
basic  positivistic  attitude, which  from my  view is  untenable, and which
seems to me to come to the same thing as Berkeley's principle, {\it esse est
percipi}. (Einstein, 1951, p. 669).  Einstein struggled until the end of his
life to  get the  observer's  knowledge back out of  physics. But he did not
succeed! Rather he  admitted that: ``It is my  opinion that the contemporary
quantum  theory...constitutes  an optimum  formulation of  the [statistical]
connections.'' (ibid. p. 87). He referred to: ``the most successful physical
theory of  our period,  viz., the  statistical  quantum theory  which, about
twenty-five years ago  took on a logically  consistent form. ... This is the
only  theory  at  present  which  permits a  unitary  grasp  of  experiences
concerning the quantum character of micro-mechanical events.'' (ibid p. 81). 

One can adopt the  cavalier attitude  that these  profound difficulties with
the  classical  conception  of nature  are  just some  temporary  retrograde
aberration in the forward march of science. Or one can imagine that there is
simply some strange  confusion that has  confounded our best minds for seven
decades, and that  their absurd  findings should be  ignored because they do
not fit our intuitions. Or one can try to say that these problems concern only
atoms  and molecules,  and not things built out  of them. In this connection
Einstein    said:  ``But  the    `macroscopic'  and    `microscopic'  are so
inter-related   that it  appears  impracticable to give up  this program [of
basing physics on  the `real'] in the `microscopic' alone.'' (ibid, p.674).

The examination of the  ``locality'' properties  entailed by the validity of
the predictions of quantum theory  that was begun by Einstein, Podolsky, and
Rosen, and was pursued by J.S. Bell,  has led to a strong conclusion (Stapp,
1997)  that  bears out  this  insight  that the  profound   deficiencies the
classical conception of  nature are not  confinable to the micro-level. This
key result will be  discussed in  section 4. But first I  discuss the reason
why, as  Mario  Bunge said:  ``The  physicist  of the  latest  generation is
operationalist  all  right, but  usually  he does not  know, and  refuses to
believe, that the original Copenhagen  interpretation --- which he thinks he
supports --- was squarely subjectivist, i.e., nonphysical.''
 
\vskip .1in

\noindent{\bf 3.  The Vulgar Copenhagen Interpretation.}
 
Let  me  call the   original    subjectivist,    knowledge-based  Copenhagen
interpretation the ``strict'' Copenhagen interpretation. It is pragmatic  in
the  sense that it  is a  practical  viewpoint  based on  human  experience,
including    sensations,  thoughts,  and  ideas.  These  encompass  both the
empirical   foundation  of our  physical theories  and the  carrier of these
theories,  and  perhaps all that  really matters to us,  since anything that
will   never   influence  any   human    experience  is, at   least  from an
anthropocentric viewpoint, of no value to us, and of uncertain realness.

Nevertheless, the prejudice of many physicists, including Einstein, is  that
the proper task of  scientists is to  try to construct a  rational theory of
nature that is  not centered  on such a  small part of the  natural world as
human experience.

The  stalwarts of  the  Copenhagen  interpretation  were not  unaware of the
appeal of that idea to  some of their colleagues,  and they had to deal with
it in  some way.  Thus one  finds  Bohr(1949)  saying,  in his  contribution
`Discussion with Einstein' to the Schilpp(1951) volume on Einstein:

\noindent ``In particular, it must be realized that---besides in the account
of the  placing  and timing  on the   instruments  forming the  experimental
arrangement---all unambiguous use of  space-time concepts in the description
of  atomic  phenomena is  confined to  the  recording of  observations which
refer to  marks on a photographic  plate or similar practically irreversible
amplification effects  like the building of a  water drop around an ion in a
cloud-chamber.'' 

\noindent and,

\noindent  ``On  the  lines of   objective  description,  it is  indeed more
appropriate  to  use the  word   phenomenon to  refer only  to  observations
obtained under  circumstances whose  description  includes an account of the
whole  experimental   arrangement. In such   terminology, the  observational
problem in quantum physics is  deprived of any special intricacy and we are,
moreover,  directly reminded  that every  atomic phenomena  is closed in the
sense that its  observation is based on  registrations obtained by
means of suitable  amplification devices with  irreversible functioning such
as, for  example,  permanent marks on  a  photographic plate,  caused by the
penetration  of  electrons  into the   emulsion. In this   connection, it is
important   to  realize  that the   quantum   mechanical  formalism  permits
well-defined applications referring only to such closed phenomena.'' 

These  are  carefully  crafted  statements. If  read  carefully  they do not
contradict the basic thesis of the strict Copenhagen interpretation that the
quantum formalism is about our observations described in plain language that
allows us to ``tell others what we have done and what we have learned.'' 

On the  other hand,  it seems  also to be  admitting  that there  really are
events occurring `out  there', which are we are  observing, but which do not
derive their realness from our observations of them.

Heisenberg (1958) says something quite similar:

\noindent ``The observation, on the other hand, enforces the description in 
space and time but breaks the determined continuity of the probability
function by changing our knowledge of the system.'' 

\noindent "Since through the observation our knowledge of the system has
changed discontinuously, its mathematical representation also has undergone
the quantum jump, and we speak of a `quantum jump' .''

\noindent ``A  real  difficulty in  understanding the  interpretation occurs
when one asks the  famous question:  But what happens  `really' in an atomic
event?''

\noindent ``If we want to describe  what happens in an atomic event, we have
to realize that the word `happens' can apply only to the observation, not to
the state of affairs between the two observations. It [ the word `happens' ]
applies to the  physical, not the  psychical act of  observation, and we may
say that the transition  from the `possible' to  the `actual' takes place as
soon as  the   interaction of  the object  with  the  measuring  device, and
therefore   with the  rest of  the  world,  has come  into  play;  it is not
connected  with the act  of  registration of the  result in  the mind of the
observer. The  discontinuous  change in the  probability  function, however,
occurs with the act of registration,  because it is the discontinuous change
in our  knowledge in the  instant of  recognition that has  its image in the
discontinuous change in the probability function.''
 
All of this is very reasonable. But it draws a sharp distinction between the
quantum formalism, which is about knowledge, and a world of real events that
are   actually   occurring  `out  there',  and  that can  be   understood as
transitions   from the  `possible' to the  `actual', closed  by irreversible
processes when the interaction  between the object and the measuring device,
and hence the rest of the world, comes into play.

Yet the extreme accuracy of detailed theoretical calculations [one part in a
hundred million in  one case] seems  to make it clear  that the mathematical
formalism must be closely connected not merely to our knowledge but also  to
what is  really  happening `out  there':  it must be  much more  than a mere
representation of our human knowledge and expectations.

I call this natural idea---that the events in the formalism correspond 
closely to real ``physical'' events out there at the devices---the Vulgar 
Copenhagen Interpretation: vulgar in the sense of common and coarse.   

This  vulgar   interpretation is I  think the  common   interpretation among
practicing     quantum   physicists:  at  this   symposium  some   important
experimentalists were,  as Mario Bunge suggested,  unwilling to believe that
the  quantum  mechanical  formalism  was about  `our  knowledge'.  And it is
coarse:  the  idea of  what  constitutes an   `irreversible'  process is not
carefully    specified,  nor is  the  precise  meaning  of  `as soon  as the
interaction with the object with the measuring device comes into play'.

My aim in this paper is to reconcile  the strict and vulgar interpretations:
i.e., to reconcile  the insight of  the founders of  quantum theory that the
mathematical  formalism of  quantum is about  knowledge  with the demand  of
Einstein that our basic physical theory be a theory of nature herself.

The  main  obstacle to a  rational   understanding of  these  matters is the
faster-than-light   action that the  quantum  formalism seems  to entail, if
interpreted at a  physical level. If  one takes  literally the idea that the
quantum event at the device constitutes a real transition from some physical
state of `possibility'  or `propensity' to a  state of `actuality' then---in
the `entangled  states' of  the kind studied  by  Schroedinger, by Einstein,
Podolsky, and  Rosen, and by Bell and  others---it would  seem that the mere
act of making a measurement in one region would, in certain cases, instantly
produce a change in the physical  propensities in some far-away region. This
apparent  faster-than-light  effect is dealt  with in the  strict Copenhagen
interpretation  by denying  that the  probability function  in the formalism
represents  anything physical: the  formalism is  asserted to represent only
our knowledge,  and our  knowledge of  far-away situations  can be instantly
changed---in  systems with   correlations---by merely  acquiring information
locally.

This fact that  the strict  Copenhagen interpretation  ``explains away'' the
apparent    violations of  the   prohibition   [suggested by  the  theory of
relativity]    of     faster-than-light   actions is  a  main  prop  of that
interpretation. 

So the essential first question in any attempt to describe nature herself is
the logical status of the claimed incompatibility of quantum theory with the
idea---from   the  theory of   relativity in   classical   physics---that no
influence can act backward in time in any frame of reference.

It is of utmost importance to progress in this field that we get this matter 
straight.

\vskip .1in
\noindent {\bf 4. Causality, Locality, and Ontology.}

David Hume cast the notion of causality into disrepute. However,  when 
one is considering the character of a putative law of evolution of a 
physical system it is possible to formulate in a mathematically clean way a 
concept of causality that is important in contemporary physical theory.

In relativistic physics, both classical and quantum mechanical,
the idea of causality is introduced in the following way:

We begin with some putative law of evolution of a physical system. 
This law is specified by picking a certain function called the Lagragian. 
A key feature of the possible Lagrangians is that one can modify them by
adding a term that corresponds to putting in an extra-force that acts only
in some small spacetime region R.

The evolution is specified by the ``law'' specified by the chosen
Lagrangian, plus boundary conditions. Let us suppose that  boundary 
condition is specified as the complete description of ``everything'' before 
some ``initial time'' $T_{in}$. The laws then determine, in principle, 
``everything'' for all later times.

In classical mechanics ``everything'' means the values of all of the
physical variables that are supposed to describe the physical system
that is being considered, which might be the entire physical universe. 

In quantum mechanics ``everything'' means all of the ``expectation values'' 
of all of the  conceivable possible physical observables, where 
``expectation value'' means a predicted average value over an 
(in principle) infinite set of instances.

To bring in the notion of causality one proceeds as follows. It is possible,
both in classical and quantum theory,  to imagine changing incrementally the
Lagrangian that specifies the law of  evolution. The change might correspond
to adding extra terms to the forces  acting on certain kinds of particles if
they  are in  some  small  spacetime  region  R that  lies  later  than time
$T_{in}$. Such a change might be regarded as being introduced whimsically by
some  outside  agent.  But,  in any  case,  one can   compare the  values of
``everything'' at times  later than time $T_{in}$  in the new modified world
(i.e., the world  controlled by the  new modified  Lagrangian) to the values
generated  from the laws  specified  by the  original  Lagrangian. If one is
dealing with  an idealized  world without  gravity, or at  least without any
distortion  of the  `flat' Minkowsky  spacetime,  then it is a  mathematical
property  of   relativistic  field   theories, both   classical and  quantum
mechanical, that ``nothing'' will be  changed outside the forward light cone
of the region R in which the Lagrangian was changed!

In other word, ``everything'' will be exactly the same in the two cases
at all points that cannot be reached from the spacetime region R without
moving faster than the speed of light.

This property of relativistic field theories is called a causality property.
The  intuition is that  this change  in the  Lagrangian can be  regarded, or
identified,  as a  ``cause'',  because it  can be  imposed  whimsically from
outside the physical  system. The mathematical  property just described says
that the effects of this  ``cause'' are confined  to its forward light cone;
i.e., to spacetime points that can be reached from the spacetime region R of
the cause  without  ever  traveling at a  speed  greater than   the speed of
light.

Relativistic field theories are formulated mathematically in such a way
that this causality property holds. This means that insofar as it is 
legitimate to imagine that human beings can ``freely choose'' [i.e., can 
act or not act upon a physical system without there being any 
cause {\it from within that physical system} of this act] to do one thing 
or another in a region $R$ [e.g., to exert or not exert a force on some 
physical particles of the system in region $R$] then ``everything'' outside 
the forward light cone of $R$ will be independent of this choice: i.e.,
``everything'' outside this forward light cone will be left unaltered 
by any change in this choice.

This relativistic causality property is a key feature of relativistic 
field theories in flat Minkowsky spacetime: it is all the causality that 
the orthodox pragmatic quantum philosophy calls for.

But notice that by ``everything'' one means, in the quantum case, 
merely the ``expectation values'', which are averages over an (in 
principle) infinite ensemble of instances.

Now, one might think that since this relativistic causality property
holds for these averages it ought to be {\it at least conceivably 
possible} that it could hold also for the individual instances.

But the amazing thing is that this is not true! It is not logically
possible to impose the no-faster-than-light condition in the individual
instances, and maintain also the validity of certain simple predictions of 
quantum theory. 

The point is this. Suppose one considers an experimental situation involving
two experimental regions that are  spacelike separated from each other. This
means that no  point in either region  can be reached  from any point in the
other without traveling faster than  the speed of light. In the first region
there is  an  experimenter who  can freely  choose to  do one  experiment or
another.  Each  of  these two   alternative   possible  experiments  has two
alternative   possible  outcomes. There is  a similar  set up in  the second
region.  Each  possible outcome is  confined to the  associated experimental
region, so   that  no  outcome of an  experiment in one  region should be
able to be   influenced by the free  choice made by the  experimenter in the
other region.

One single instance is considered,  but with the two free choices of the two
experimenters  being  treated  as two free   variables. Thus the  one single
instance under  consideration, uniquely fixed at  all times earlier than the
earliest time in either of the two experimental regions, will go into one or
another of altogether $(2\times 2)=$ four alternative possible evolutions of
this system, depending  on which of the two  alternative possible choices is
made by each of the two  experimenters. There can then be further branchings
that are specified by which of the  possible   outcomes  nature  selects for
whichever experiments are performed.

The  particular  experimental  details can be  arranged so  that the assumed
validity   of the    predictions of   quantum  theory  for  that  particular
arrangement  entails  the {\it   nonvalidity} of at  least one  of the three
following locality conditions:

{\bf LOC1:} It is possible to impose the following condition: If in each of
the two regions the first of the two possible experiments were to be 
performed, and a certain result $r$ appeared in the first region then
if this very same experiment were to be performed in the first region
then this same result $r$ would appear there even if the experimenter in 
the second region were to elect at the last moment to do the other
measurement.  

The rationale for this  locality condition is  that a free choice of what to
do in one place cannot---relativity theory leads us to believe--- affect, at
a speed  faster than the  speed of  light, what  occurs  elsewhere: making a
different choice in one region should  not be able to force what appears (at
the  macroscopic,  observable level)  in the  other region to  be different.
Indeed,  in some  frame of  reference the  outcome in  the first  region has
already occurred before the experimenter in the second region makes his free
choice of  which experiment  he will  perform. But,  according to ideas from
relativity theory, what  someone has already seen  and recorded here at some
earlier  time cannot be  disturbed by  what a  faraway  experimenter  freely
chooses to do at some later time.

Notice that LOC1 requires only that it be possible to impose this condition.
The point is that only  one of the two possible  experiments can actually be
performed in the second region, and  hence nature herself will make only one
choice.   So  what  would   actually  appear  in  the  first  region  if the
experimenter in  the other (far away)  region were (at  some future time) to
make  a  different  choice in not  physically  well defined.  Thus this is a
theoretical   investigation: the  question is whether  the predictions of QT
are compatible  with the notion that  nature evolves in such a way that what
one  observer sees  and  records in  the past  can be  imagined  to be fixed
independently of what another person will freely choose to do in the future.

{\bf LOC2:} Suppose, under the  condition that the first of the two possible
measurements  were to be  performed in the  first region  (with no condition
imposed on what the  outcome there is) that one  can prove from LOC1 and the
predictions of  quantum theory, the  truth of a  statement $S$ that pertains
exclusively  to  what   experimenters can   observe under  various  possible
conditions  of their own  making in  the second  region. Then  this locality
condition asserts  that it is  logically possible to  demand that $S$ remain
true under the  condition that the  experimenter in the  first region freely
chooses (say in the  future) to perform there,  instead, the second possible
measurement.
 
The  rationale  is  that,  according to  certain  ideas  from the  theory of
relativity, the truth of a statement that pertains to macroscopic conditions
that refer  exclusively to one  space-time region should  not depend on what
someone far away freely chooses to do later.

{\bf LOC3} This is another form of LOC1: Altering the free choice in R
leaves any outcome in L undisturbed. [See Stapp, 1997]

The validity of the predictions of  quantum theory in correlation situations
like this  are being  regularly  borne  out. (...Most  recently  in a highly
publicized  experiment using the  Swiss telephone  company optical fibers to
connect   experimental regions  that were  14 km  apart, with  the intent of
making   important  practical   applications.)  Thus it  can, I  believe, be
confidently assumed that the pertinent quantum predictions are valid. But in
that case one of the ``locality conditions'' described above {\it must} fail. 

Before drawing any conclusions one  must consider the impact or significance
of the assumption that the  experimenters' choices can be treated as  ``free
variables''.

It is  part of  the  orthodox  quantum  philosophy that  the  experimenters'
choices can and  should be considered  to stand outside  the physical system
that is being examined.  Bohr and Heisenberg  argued that biological systems
in  general lie outside  the domain covered by  the pragmatic framework. But
in  any  case, one  thing  is  certain: the  beautiful  and  elegant quantum
formalism  is naturally suited to the  idea that it represents a system that
is part  of a bigger system that can  extract information from it, where the
nature of the  information being extracted from  the subsystem is controlled
by things outside that subsystem, namely the observer and his instruments of
observation.

But even at a more intuititive level it seems that the decision-making 
process of human experimenters are so complex and delicate, and so 
insulate-able in principle, prior to the time of the examination, 
from the system that they are about to examine, as to render their choices 
as to what to look for effectively free, under appropriate conditions of
isolation, from any influence upon them {\it by the system they are about to 
examine}. So it would seem to be safe, under appropriate conditions of
prior isolation, to treat these choices {\it as if they were free from such 
influences} even in a strictly deterministic universe. 

In a quantum universe this move is even more reasonable, because the 
choices could be governed by a quantum process, such as the decay of a 
radio-active nucleus. Within the quantum theoretical framework each
such decay appears as a spontaneous random event. It is free of any 
``physical'' cause, where ``physical'' means something that is part of 
the physical world as that world is described by the physical theory.
Thus within both the deterministic and stochastic contexts it seems
reasonable to treat the choices to be made by the experimenters as if
they were free, in the sense of not being influenced by the physical 
properties of the system that is about to be examined. 

One caveat. The arguments require that meaning be given to  
a condition such as: ``If the experimenter in region one performs 
experiment one, and the outcome that occurs there is outcome one''.
This condition is nonsensical in the Everett many-minds interpretation, 
because {\it every} outcome occurs. I have excluded that interpretation 
from consideration on other grounds, which are described in section 5.

The apparent failure of the locality condition has three important 
consequences:

1. It gives a  solid  basis for the  conclusion of  the  founders of quantum
   theory   that  no   return  to  the   notions  of   classical   mechanics
   (relativistic  field theory) is  possible: the  invalid locality property
   certainly holds in relativistic classical mechanics.

2. It  makes   reasonable  the   attempt to    ontologicalize  the  orthodox
   interpretation. It had formerly been believed that this was a nonsensical
   thing  to   try,  because      ontologicalization    immediately  entails
   faster-than-light   transfer of  information on  the  individual-instance
   level.  Such  transfers had seemed  unacceptable, but  are now seen to be
   unavoidable  even in a  very general framework  that maintains merely the
   validity of  the {\it  predictions} of quantum  theory, and the idea that
   the  experimenters' choices can be considered to be ``free'', in the weak
   sense discussed above.

3. Because   the  nonlocal   effects  enter into   orthodox  quantum  theory
   specifically  in   connection with  the  entry of our   knowings into the
   dynamics   there  is  prima  facie  evidence  that  our  knowings  may be
   associated with  the nonlocal  aspect of nature. It  is worth noting that
   these  effects are  not  confined to a  microscopic  scale:  in the Swiss
   experiment  the effect in  question  extended over a  separation of 14km.
   And,  according to quantum  theory, the  effect does not  fall off at all
   with distance. In my  proposal each of our  knowings is associated with a
   brain event that involves, as a unit, a pattern of brain (e.g., neuronal)
   activity that  may extend  over a large  part of the  brain. The collapse
   actualizes   this  whole   pattern,  and the   associated   knowing is an
   expression of the functional properties of this pattern. 
 
Once the  reality is  recognized to  be  knowledge, rather  than substantive
matter,  the  nonlocal   connections seem  less   problemmatic:  nothing but
knowledge about far-away knowings is changed by nearby knowings.

\vskip .1in

\noindent {\bf 5. All Roads Lead to Solvay 1927.}

The Solvay conference of 1927 marks the birth of (coherently formulated)
quantum theory. Two of the many important papers delivered there stand out.

Born and Heisenberg presented a paper on the mathematical formalism and
proclaimed that the essential features of the formalism were complete and 
not subject to further revision.

Dirac gave a paper on the interpretation, and claimed
that ``the wave function represents our knowledge of the system, and the 
reduced wave packets our more precise knowledge after measurement.''

These two parts, the mathematical formalism and its interpretation 
in terms of knowledge, meshed perfectly: that was the logical basis of the
Copenhagen interpretation.

This was an epic  event in the  history of human  thought. Since the time of
the  ancient  Greeks the  central  problem in   understanding the  nature of
reality, and our role in it, had been the puzzling separation of nature into
two seemingly  very different  parts, mind  and matter.  This had led to the
divergent approaches of idealism and  materialism. According to the precepts
of idealism our ideas, thought, sensations, and other experiential realities
should be  taken as basic.  But then the  mathematical  structure carried by
matter was difficult to fathom in any natural way. Materialism, on the other
hand, claimed that  matter was basic. But, if one  started with matter  then
it was  difficult to  understand how  something like your  experience of the
redness  of  a red   apple  could be    constructed out  of  it, or  why the
experiential   aspect  of  reality  should  exist at  all  if, as  classical
mechanics  avers,  the  material aspect is  dynamically  complete by itself.
There seemed to be no rationally coherent way to comprehend the relationship
between our experiences of the reality that exists outside our thoughts, and
the  nonexperiential-type material  substance that the  external reality was
claimed to be made of.

Yet at the  Solvay meeting,  physicists, of  all people,  had come up with a
perfect  blending,  based on  empirical evidence, in  which the mathematical
structure needed to  account for all of the  empirical regularities formerly
ascribed  to  substantive matter, was  present without  there being anything
like   substantive  matter: the   mathematical  structure was a  property of
knowings!

What an exhilerating moment it must  have been. Driven simply by the need to
understand in a rational  way the empirical facts  that nature had presented
to us,  scientists  had  been led to  a  marvelous  resolution of  this most
fundamental of all philosophical  problems. It was a tremendous achievement.
Now, seventy years  later, we are  able to gather here  at the X-th Max Born
Symposium to  celebrate the  unbroken record  of successes  of that profound
discovery, and to hear about its important new triumphs.

So now, the end of our Symposium, I take this opportunity to review briefly
some of its highlights from the perspective of the Solvay breakthough.

Probably the most  exciting reports  were from  experimentalists who are now
performing  experiments that could  only be imagined  seventy years ago. Yet
the  thinking of the  founders of  quantum theory  did involve  ``gedanken''
experiments  designed to  confirm the  rational coherency  of the framework.
Today these ``thought'' experiments  involving preparations and measurements
on small numbers of individual atoms  are being carried out, and the results
invariably  confirm all  of the  ``quantum  weirdness'' that  the Copenhagen
interpretation predicted. 

But do these successes really confirm the radical ideas of Solvay 1927? Time
has eroded the message of Solvay to the extent that the scientist performing
the experiments hardly  recognize the Solvay  insights in the interpretation
of their work, though  they give lip service to  it. One must probe into the
rational  foundations of the subject  to see the import  of their results on
this deep question.

I cite first the report  of Omnes. There had been  hope that some way around
the Copenhagen  interpretation would emerge from  the studies of decoherence
and consistent  histories that have  been so vigorously  pursued of late. No
one has  pursued these  ideas more  deeply than  Omnes. His  verdict is that
these methods amount to ``the  Copenhagen interpretation `done right' ''. He
said similar things in his book (Omnes, 1994). And such prominent proponents
of ``decoherence'' as  Zurek(1986) and Joos(1986)  have said similar things:
Zurek   concluded that  the  study of   decoherence  ``constitutes  a useful
addition to the  Copenhagen ...a clue pointing at  a still more satisfactory
resolution of the  measurement problem...a hint  about how to proceed rather
than the means  to  settle the matter quickly.''  Joos asks at the beginning
of his article ``Is there some way, at least a hint, how to understand... ''
and at the end  says ``one may hope  that these  superselection rules can be
helpful in developing new ideas ..[about].. measurement processes.'' So they
both  stressed that decoherence effects do not resolve the deep problems. 

Indeed,  decoherence is rather the  {\it cause} of the  problem: decoherence
effects  make it  virtually   impossible to  empirically   determine whether
quantum collapses are  occurring outside our  brains or not. It is precisely
{\it  because} of  decoherence  effects  that we  cannot tell,  empirically,
whether or not  collapses  actually do occur  ``when the  interaction of the
object with  the measuring  device, and  hence the rest of  the world, comes
into play''.

The   decoherence-consistent-histories approach had  originally been pursued
within  the   Everett   framework,  and  indeed  was   sometimes  called the
`post-Everett'  approach  to stress  that it was  being  pursued within that
framework, rather than the Copenhagen  framework, which it sought to unseat.
But Omnes put his finger  on the fatal flaw in  the Everett approach when he
said that it did not  explain the transition from  ``and'' to ``or''. In the
evolving   wave   function  of  Everett  the   various   branches do  evolve
independently,  and  hence  might  naturally be  imagined to  have different
``minds'' associated with them, as Everett suggests. But these branches, and
the minds  that are  imagined to be  properties  of these  branches, are all
simultaneously present. Hence there  is no way to give meaning to the notion
that one mind is far more likely to  be present at some finite time than the
others. It is like waves on a pond: the big waves and the small ones are all
present simultaneously.  So one needs something  else, perhaps like a surfer
that will be pushed into one branch  or the other, to define the ``or'' that
is  logically  needed to  define  the  notion of  the   probabilities of the
different  ``alternatives''. Yet the  Everett  interpretation allows nothing
else  besides the wave  function and  its  properties. So all  the minds are
simultaneously   present  because all the   corresponding  properties of the
various branches are simultaneously present.

The idea of the surfer  being pushed by the wave  is exactly the idea behind
the model of David Bohm  that was so ably  expounded here by D. Duerr and by
F.  Faisal.  But  the  model has  not  been   consistently  extended  to the
relativistic case of quantum  electrodynamics, or to quantum chromodynamics,
which are our premiere quantum theories. 

The model has  other  unpleasant features.  One is the  problem of the empty
branches. Each  time a ``good  measurement'' is  performed the wave function
must separate into  different ``branches''. These  branches are parts of the
wave    function  such  that  the  full   wave   function  is a  sum  (i.e.,
superposition)  of  these  branches, and  each branch  is  nonzero only in a
region (of the 3n-dimensional space in which these wave functions live) that
overlaps none of the other regions. Here n is the number of particles in the
universe.

If two branches separate then the `surfer' (which in the Bohm model would be
the entire classically described  physical world) must end up in just one of
these branches. But all the other branches (which are regarded as physically
real)  must continue  to evolve  for all  eternity  without ever  having any
effect upon the `surfer', which is the only part of reality that is directly
connected to  human  experience, according  to the model.  This seems wildly
extravagant! If the surfer is the  important thing then the effort of nature
to continue to evolve these  ineffectual branches for  all eternity seems to
be a gigantic waste of effort. But if  the surfer is  not important then why
is this tiny part  of reality there  at all? It does  nothing but get pushed
around. 

There is a perhaps bigger problem  with the initial conditions. The model is
predicated on the  premise that the single real  classical world is a random
element  in  a   statistical   ensemble of    possibilities.  The  idea of a
statistical   ensemble  makes good  sense when  we have  the  possibility of
repeated preparations of similar  situations. But when we are speaking about
the entire universe it does not seem  to make sense to speak of a particular
statistical  ensemble of  universes  with some  particular  density (weight)
function if only  one of them is ever  created. Or are  we supposed to think
that a whole ensemble of real  classical worlds is created, and that ``our''
real  world is just  one of  them? That  would seem  to be the  more natural
interpretation.  But I asked David  Bohm about that,  many years ago, and he
insisted that there was, according to his thinking, only one universe.

Bohm was stimulated to construct his model by conversations with Einstein.
Yet Einstein rejected the model, calling it ``too cheap''. 

I asked Bohm what he thought about Einstein's evaluation, and he said he
completely agreed.

Indeed, at the end of  his book with Hiley about  his model, after finishing
the part describing the model, he  added two chapters about going beyond the
model. He motivated  those chapters by references  to the efforts that I was
making,  and  that  Gell-mann  and  Hartle  were  making,  to go  beyond the
Copenhagen     interpretation.   Gell-mann  and  Hartle  were   pursuing the
decoherence-consistent-histories   approach  mentioned  above, which has led
back to Solvay, and I had proposed a  theory of events. The events were real
collapses of a wave function that was considered to be ontologically real.

This  brings me to the  talk of  Rudolf Haag.  Haag described  his theory of
events, and  mentioned  that it still  needed  twenty years of  work. In his
written account Haag(1996) mentions that I had proposed essentially the same
theory in the  seventies, some twenty  years ago (Stapp,  1975, 1977, 1979).
{\it My} twenty years of work on this idea has lead back to Solvay 1927. The
problem is always the same: if one  wants to make natural use of what nature
has told us, namely that the  beautiful mathematical formalism works to high
precision,  then one is led  to ascribe to  that formalism  some ontological
reality. But  then the  condition for the  collapses must  be spelled out in
detail. 

It  is  natural  for   physicists to  try  to  find  {\it  purely  physical}
conditions.  But in the end there are no adequate natural conditions of this
kind: the  possibilities  are all unnatural and  ad hoc. Von Neumann said it
all when he  showed, back in the  thirties, that one could push the boundary
between  the   world   described by  the  quantum   formalism and  the world
described  in terms   our  classical  concepts all  the way to  the boundary
between  brain and  mind   without  disrupting  the  predictions  of quantum
theory, and noted that there is no  other natural place to put the boundary,
without    disrupting the   integrity of  the  theory.  In  fact, it  is, in
principle, {\it only} if  one pushes the boundary  all way to the brain-mind
interface that  one obtains,  strictly, the  prediction of  orthodox quantum
theory: otherwise  there are rogue  collapses  that are  not associated with
knowings.

Of course,  pushing the  boundary all the  way to mind  brings mind into our
theory of nature. But why on earth should we try to keep mind out--- bottled
up,  ignored, and  isolated  from the  physical  world---when we  know it is
present, and seemingly  efficacious, particularly  when the intense struggle
of  physicists  to  find a  rational  way of   accounting  for the  observed
phenomena led them to the conclusion that the theory of physical reality has
the form of a theory about knowings,  not the form of a theory about matter.
Our aim should be not to  bring back moribund  matter, which we are well rid
of, but to learn how  better to understand  knowings, within the mathmatical
framework provided for them by the  quantum formalism. 

\vskip .1in 
\noindent {\bf 6. The Two Quantum Processes.}

There  have  been many  attempts  by  physicists  to `get  mind  back out of
physics': i.e., to reverse the  contamination of physics brought in by Bohr,
Heisenberg,     Dirac,   Pauli  and   company  in  1927.  I   believe  those
decontamination efforts  have failed, even though  I myself have worked hard
to achieve it.  So I am  taking here the  other tack, and  trying to build a
coherent ontology around  the orthodox ideas. In  particular, I am accepting
as basic the idea that there are knowings, and that each such knowing occurs
in conjunction  with a  collapse of the wave  function that  reduces it to a
form concordant with that knowing. I assume that knowings are not associated
exclusively   with  human   body/brains. But  I  shall  focus here  on these
particular kinds of knowings because these are the ones we know most about. 

A fundamental fact of  orthodox quantum theory is  that the evolution of the
state of the physical  system between the  collapse events is mathematically
very   different from  the  evolution  of  this  state  associated  with the
collapses: the former are ``unitary''  and ``local'', whereas the latter are
neither.

The ``unitarity''  property means several things.  On the one hand, it means
that the   evolution is in  some  sense no  change at  all: the  internal or
intrinsic structure of the state is unaltered. One can imagine that only the
`mode of description' of the state is changed, not the state itself. Indeed,
that point of view is very often adopted in quantum theory, and is the one I
shall adopt here. (See the next section.)

The ``unitarity'' property also means that the transformation operator that 
changes the state at an earlier time to the state at a later time does not 
depend on that initial (or final) state: there is, in this sense, 
in connection with the unitary part of the process of evolution, 
{\it no self reference!}

According to the orthodox interpretation, there is no experiential reality 
associated with the unitary part of the evolution, which is the part between 
the observations: there is no essential change, and no self reference, and 
hence, reasonably enough, no experience.

Experiences are associated only with  the nonunitary parts of the evolution:
the  part  associated  with   observations.  For  that part  there  {\it is}
essential   change, and  the  transformation  operator  analogous to the one
defined for  the unitary case would  depend on the state upon which it acts.
Thus there   would be, in  this sense,  self-reference.  This self reference
(nonlinearity) plays a key role in the dynamics associated with observation.
It is a special  kind of self reference that has no counterpart in classical
mechanics.

In the  classical  approximation to the  quantum dynamics  only the  unitary
part of the  dynamical  evolution  survives. So  from a  quantum  mechanical
point  of  view it  would  be   nonsensical to  look  for  mind in a  system
described by  classical physics. For  classical physics  is the result of an
approximation to  the full dynamical  process of nature  that eliminates the
part  of that process  that orthodox quantum  theory says is associated with
our  experiences. 

\vskip .1in

\noindent {\bf 7. The Two Times: Process Time and Mathematical Time}

The distinctions between the two processes described above is central to
this work. It can be clarified, and made more vivid, by explaining how 
these two processes can be considered to take place in two different times.

In quantum theory there are two very different kinds of mathematical
objects: vectors and operators. Operators operate on vectors:
the action of an operator on a vector changes it to another (generally
different) vector.

Given an operator, and a vector that represents a state of a physical
system (perhaps the entire universe), a number is formed by first letting
the operator act on the vector, and then multipling the resulting vector
by the (complex conjugate of the) original vector. This number is called
the ``expectation value of the operator in the state represented by that
vector''.

Modern field theories are generally expressed in the so-called Heisenberg 
picture (rather than the so-called Schroedinger picture). I shall follow 
that practice.

In ordinary relativistic quantum field theory each spacetime point has a 
collection of associated operators. (I gloss over some technicalities
that are not important in the present context.)

Consider the collection of operators $C(t)$ formed by taking all of the
operator associated with all of the spacetime points that lie at 
fixed time $t$.

This set $C(t)$ is ``complete'' in  the sense that the expectation values of
all the  operators of $C(t)$  in a state $S$  determine all  the expectation
values of the all the operators in $C(t')$ in the state $S$, for every  time
$t'$. The  operators in  $C(t)$ are related  to those in  $C(t')$ by a  {\it
unitary} transformation. Whether one represents the state $S$ by  giving the
expectation values in  this state of all the  operators in $C(t)$, or of all
the operators in $C(t')$, is very  much like choosing to  use one coordinate
system or  another to  describe a  given  situation: it is  just a matter of
viewpoint.  The  unitary   transformation  that  relates  the  collection of
operators $C(t)$ to the collection of  operators  $C(t')$ is essentially the
unitary  transformation associated  with the  Schroedinger-directed temporal
evolution.  It  is in  this  sense  that the   unitary   transformation that
generates evolution in the  ``mathematical time'' $t$ is relatively trivial.
It is deterministic,   continuous, invertible,  and independent of the state
$S$ of the physical  system upon which the operators act.

But giving  the complete set  of all the  operators  associated with all the
points in  spacetime says nothing at  all about the  evolution of the state!
Saying everything that can be said about the operators themselves, and about
evolution via the  unitary part of  the transformation  has merely fixed the
mode  of   description, and  the   connections  between  different  modes of
description. It has not  said anything about the  all-important evolution of
the state.

The state undergoes a sequence of abrupt jumps:
 $$...S_i \longrightarrow S_{i+1}  \longrightarrow S_{i+2}...  .$$
 
The situation can be displayed graphically by imagining that $i$ is the
imaginary part of the complex time $t$: the evolution proceeds at constant 
imaginary part of $t$ equal $i$, and at constant $S_i$, with the
real part of $t$ increasing until it reaches a certain `jump time' $t_i$, 
whereupon there is an abrupt quantum jump to a new constant state $S_{i+1}$,
and a new constant imaginary part of $t$ equal to $i+1$, and the evolution
then again proceeds with increasing real part of $t$ until the next `jump
value' $t_{i+1}$ is reached, and then there is another jump up to a new
value, $i+2$, of the imaginary part of $t$. Thus the full process is 
represented in complex time as a line having the shape of a flight of steps.
The horizontal segments where the real part of time is increasing
represent the trivial unitary parts of the process, which correspond
merely to changing the viewpoint, or mode of description, with the state
remaining fixed, and with no associated experience. The vertical segments
correspond to increases in `process time'. These are the parts associated
with experience. (This identification of the vertical axis with imaginary
time is purely pedagogical)

The  present endeavour  is to begin  to fill in  the details  of the process
associated  with the  increases in  the vertical  coordinate,  process time,
which is the time  associated with  the nontrivial part  of the evolutionary
process,  and with  experience. The final phase  of each vertical segment is
the fixing  of a  new knowing. But  some process in  Nature must bring about
this  particular  fixing:  this process is  represented by  motion along the
associated vertical segment. 

\vskip .1in

\noindent {\bf 8. Quantum Ontology}

What is the connection between the our experiences and the physicists'
theoretical description of the physical world?

The materialist position is that each experience {\it is} some aspect of the 
matter from which the physicists say the world is built.

But the physical world certainly is not built out of the substantive matter 
that was postulate to exist by classical mechanics. Such stuff simply does
not exist, hence our experiences cannot be built out of it.

The quantum analog of physical reality, namely the quantum state S of the 
universe, is more like information and ideas than like the matter of classical
physics: it consist of accumulated knowledge. It changes when human knowledge
changes, and is tied to intentionality, as I shall explain presently.

Orthodox classical mechanics is naturally complete in itself: 
the physical world represented in it is dynamically complete,
and there is no hint within its structure of the existence of anything else.

Orthodox   quantum   mechanics is  just the   opposite: the   physical world
represented by it is not dynamically  complete. There is a manifest need for
a process that is not represented within the orthodox description. 

In  orthodox quantum  mechanics the  basic  realities are our  knowings. The
dynamics of the physical world represented in the orthodox quantum formalism
is not   internally  complete  because  there is,  in  connection  with each
knowing,  a  collapse  process  that  appears in  the  orthodox  theory as a
``random choice''  between  alternative possibilities:  contemporary quantum
theory provides no  description of  the process that  selects the particular
knowing that actually occurs.

This  collapse   process, which  is  implemented  by a   nonunitary/nonlocal
transformation, must specify two things that the contemporary machinery of 
quantum theory does specify:

\noindent 1. It must specify an experience E, associated with a corresponding 
projection operator P(E), such that the question is put to Nature: 
``Does E occur?''

\noindent 2. It must then select either the answer `yes', and accordingly
change the current state (i.e., density matrix) S to the state PSP, or 
select the answer `no', and accordingly replace S by (1-P)S(1-P). The 
probability of answering `yes' is Trace PSP/TraceS; the probability of 
answering `no' is Trace (1-P)S(1-P)/Trace S.

In the orthodox pragmatic interpretation the step 1 is achieved by a human
experimenter's putting in place a device whose observed response will 
determine whether the system that is being examined has a certain property
specified by P(E): the occurrence of experience E will confirm, basically
on the basis of past experience, that future experiences will be likely to
conform to the answer ``Yes, the system has property P(E).'' 

According to the orthodox viewpoint, the experimenter stands outside the
quantum system being examined, and the device is regarded as an extension
of himself. 

Step 2 is then achieved by appeal to a random selection process
that picks the answer `Yes'  or `No' in accordance with a statistical 
rule. This selection process (also) is not represented within the orthodox
Hilbert space description.

How can these two steps be comprehended in a rational, minimalistic, 
naturalistic way?

\vskip .1in

\noindent {\bf 9. Von Neumann's Process I.}

The first step in the nonunitary process is what von Neumann called
Process I, in contrast to his Process II, which is the normal unitary
evolution.

Process I consists of  ``posing the next  question". We can suppose that the
possible answers  are $Yes$ or $No$.  Nature will then  answer the question.
The  crucial  requirement  is that  the answer  $Yes$  must be  recognizably
different from  the answer  $No$, which  includes no  recognizable answer at
all.

In practice a human being creates the conditions for Process I, and it is he
who recognizes the positive response: this recognition is a knowing.

For example, the observer may know that he is seeing the pointer on the
device---that he himself has set in plac---resting definitely between the 
numbers 6 and 7 on the dial. This is a complex thing that he knows. 
But knowings can be known, at least in part, by later knowings. This is 
the sort of knowing that science is built upon. Of course, all one can 
really know is that one's experiences are of a certain kind, not that there 
really is a pointer out there. So we expect the knowings to correspond in 
some way to a brain activity of some sort, which under normal circumstances 
would be an effect of something going on outside the brain.

Von  Neumann  accepts  the  statistical   character of  the  theory, and his
Process I  is  statistical in  character: his  Process I  covers merely  the
posing of the question,  and the assignment of a  statistical weight to each
of the  recognizably  different  alternative  possible  answers. It does not
cover the subsequent process whereby Nature delivers an answer.

My basic  commitment here is to  accept the quantum  principles as they are,
rather than to  invent new principles  that would allow  us to  exclude mind
from  Nature's   dynamics. So  I  accept  here,   ontologically as  well  as
pragmatically,  that  {\it the  possibilities  singled out in  Process I are
defined by different `possible knowings'}.
  
Two important features of the von Neumann Process I are:

1) It produces an abrupt  increase in entropy. If  the state of the universe
   prior to the process is well defined, so that the entropy (with no coarse
   graining)   is  zero,  then  if,  for  example,  the   Process I  gives a
   statistical  mixture with  $50\%\ Yes$ and  $50\%\ No$,  the entropy will
   jump to $ln 2$.

2) It is quasi-local. There will be nonlocal aspects extending over the size
   of the examined  system, but no  long-range nonlocal  effects of the kind
   mentioned  in  section  3. That  is,  there  will be,  for the  Process I
   associated with a human knowings, brain-sized nonlocal effects associated
   with defining the question, but no nonlocal effects extending outside the
   body/brain. Thus Process I is, for human knowings, a human process, not a
   global  one.  [Technically,  the reason  that there is  no effect  on 
   far-away
   systems is that such an  effect is computed by  performing a `trace' over
   the degrees of freedom  of the nearby system  (e.g., the brain/body), but
   von Neumann's  Process I is  achieved by dropping out  interference terms
   between the alternative possible  answers, and that operation leaves this
   trace unaltered.]

Process  I lies  at the  root of   measurement  and  mind-body  problems. In
approaches  that try to  explain  Process I in  purely  physical terms, with
knowings  not   mentioned,  but  rather  forced to  follow  from  physically
characterized processes,  the answers tend to  assert either that:\\ 1), the
wave function  of a particle  occasionally  just  spontaneously reduces to a
wave function that is essentially zero except over a small region, or that\\
2),  what is  not   measurable in   practice  (i.e.,  via  some  practicable
procedure) does  not exist in  principle: if it is  impractical to detect an
interference term them it does not exist.

This latter sort of rule is certainly justified in a pragmatic approach. But
most physicists have been reluctant  to accept such rules at the ontological
level. Hence the pragmatic approach has won by default.

From the  present  standpoint, however, the  basic  principle is that Nature
responds only to questions that are first posed, and whose answers are 
possible knowings, or are things of the same general ontological type as 
possible knowings. [The needed generalization will be discussed later, after 
the knowings themselves have been discussed.]  

But the important immediate point is that the quantum dynamics is organized
so as to put knowings, and their possible generalizations, into the central 
position.

All such knowings contribute to the  general self knowledge of the universe,
which is represented by the (Hilbert-space) state $S$ of the universe.

\vskip .1in
\noindent {\bf 10. Origin of the Statisical Rules}
\vskip .1in

Without  loss of  generality we can  suppose that  each posed  question is a
single question answered  with a single reply,   {\it Yes} or {\it No}. Then
the usual  (density  matrix)  formalism allows  the reduction  process to be
formalized in the following way. The state of the universe is represented by
the  density  matrix  (operator) $S$.  The  question is   represented by the
projection   operator $P$:  $P^2 =  P$. Then  the von  Neumann  Process I is
represented by $$ S\equiv  [PSP+(1-P)S(1-P)+PS(1-P)+(1-P)SP] \longrightarrow
PSP +  (1-P)S(1-P). $$ The  subsequent  completion of the  reduction is then
represented   by $$  [PSP +   (1-P)S(1-P)]    \longrightarrow PSP  \mbox{ or
}(1-P)S(1-P) $$ where the fractions  of the instances giving the two results
are: $$ (Tr PSP)/(Tr PSP  + Tr (1-P)S(1-P))  \mbox{\ for\ } PSP $$ \noindent
and $$ (Tr (1-P)S(1-P))/(Tr PSP + Tr(1-P)S(1-P)) \mbox{\ for\ } (1-P)S(1-P).
$$ Here Tr represents the trace operation, which instructs one to sum up the
diagonal  elements  $<i|M|i>$ of the  matrix  $<j|M|i>$ that  represents the
operator, for some  complete orthonormal set of  states $|i>$. [The value of
the trace does not depend upon which  complete orthonormal set is used, and,
for any two  (bounded)  operators $A$ and  $B$, $Tr AB = Tr  BA$. Using this
property, and $P^2 = P$, one sees  that the denominator in the two equations
just given reduces to $Tr S$. A  partial trace is given by the same formula,
but with the vectors $|i>$ now forming a complete orthonormal basis for 
{\ part} of the full system]

I believe  it is  perfectly  acceptable to  introduce an  unexplained random
choice or selection in a  pragmatically formulated theory. But in a rational
ontological  approach there  must be some  sufficient cause  or reason for a
selection to pick out $Yes$ rather  than $No$, or vice versa. In view of the
manifestly nonlocal character of the  reduction process, there  is, however,
no reason for this selection to be determined locally.

Quantum theory does not specify what this selection process is, and I do not
try to do so. But given our ignorance of what this process is, it is highly 
plausible that it should give statistical results in accord with the 
rules specified above. The reason is this.

If the selection  process depends in some unknown  way on things outside the
system being examined then the  fractions ought to be invariant under a huge
class  of  unitary   transformations $U$  of the  state  $S$ that  leave $P$
invariant,   for   these    transformations  are   essentially  the  trivial
rearrangements of the distant features of the universe:

$$
S\longrightarrow USU^{-1} \ \ \ \ \ \ \ U^{-1}PU = P.
$$ 

Since the statistical description after the Process I has occurred
is essentially similar to the classical statistical description one
should expect $S$ and $P$ (or $(1-P)$) to enter linearly. But the trace
formulas are the only possibilities that satisfy these conditions, for all
$U$ that leave $P$ invariant.

The point here is only that if the actual selection process depends in a 
complicated and unknown way on distant uncontrolled properties of $S$ then 
the long-term averages should not be sensitive to basically trivial 
rearrangements made far away.

This  assumption is  quite  analogous to  the  assumption made  in classical
statistical analysis---which has a  deterministic underpinning---that in the
absence of  information  about the  full details  one should  integrate over
phase space {\it without  any weighting factor  other than precisely one} in
those  degrees of  freedom  about  which one  has no   information. Thus the
quantum statistical  rules need not be regarded  as some mysterious property
of nature to have unanalysable {\it tendencies} to make sudden random jumps:
it is rational to  suppose, within an  ontological  setting, that there is a
causal, though certainly  nonlocal, underpinning  to these choices, but that
we do  not yet  know  anything about  it, and  hence our   ignorance must be
expressed by the uniquely appropriate  averaging over the degrees of freedom
about which we have no knowledge.

The  effective   randomness of  Nature's  answers  does  not  render the our
knowings nonefficacious.  Our knowings can enter  the dynamics in a strongly
controlling way {\it  through the choice of the  questions}, even though the
answers to  these  questions are  effectively  random. The  formation of the
questions, in  Process I, is human  based, even though  the selection of the
{\it answers} is presumably global. This will be discussed presently.

The theory is  naturalistic in that, although  there are knowings, there are
no soul-like  experiencers: each human stream of  consciousness belongs to a
human body/brain, which provides the structure that links the experiences of
that stream tightly  together.

\vskip .1in
\noindent {\bf 11. Brains and Experiences.}
\vskip .1in

The dynamics of the theory is organized around the collection of operators 
P(E) that connect experiences E to their effects on the state S of the 
universe. I describe here my conception of this connection, and of the
dynamical differences between the quantum version of this connection
and its classical analog.

Each experience is supposed to be one gestalt that, like a percept,
``comes totally or not at all'', in the words of Wm. James (1987. p. 1061). 
This experience is part of a sequence whose elements are, according to James, 
linked together in two ways: each consists of a fringe that changes only 
very slowly from one experience to the next, and a focal part that changes 
more rapidly. The fringe provides the stable contextual framework. It is
the background that provides both the contextual setting, within which 
the foreground is set, and the experience of a persisting historical 
self that provides both the backdrop for the focal part and the carrier
of longer term motivations. The focal part has a sequence of temporally 
displaced components that, like the rows of a marching band that are 
currently in front of the viewing stand, consists of some that are just 
coming into consciousness, some that are at the center, and some that are 
fading out. The occurrence together, in each instantaneous experience, 
of this sequence of temporal components is what allows comparisons to be
made within a conscious experience. Judgments about  courses of events
can be parts of an experiences. The experiences are organized in the first
instance, around experiences of the person's body  in the context of his 
environment, and later also around abstractions from those primitive 
elements. These matters are discussed in more detail in chapter VI
of my book (Stapp ,1993).

Each experience normally has a feel that includes an experience of a 
prolongation of the current sequence of temporal components: this 
prolongation will normally be a prolongation that is, on the basis of 
past experience, likely to be imbedded in the ``current sequence of temporal 
components'' of some later experience in the linked sequence of experiences. 

Each experience E induces a change of the state of the universe S--$>$ PSP. 
This change will, I believe, for reasons I will describe presently, be a 
specification of the classical part (see below) of the electro-magnetic field 
within the brain of the person. This specification will fix the activities
of the brain in such a way as to produce a coordinated activity that will 
generally produce, via a causal chain in the physical world (i.e., via
the causal evolution specified by the Schroedinger or Heisenberg equations 
of motion) the potentialities for the next experience, $E'$.
That causal chain may pass, via the motor cortex, to muscle 
action, to effects on the environment, to effects on sensors, to effects 
on the brain, and finally to a set of potentialities for various possible
prolongations of the current sequence of temporal components.

Then a selection must be made: one of the potential experiences will become
actual. 

But this description glosses over an essential basic problem: How do the
possible experiences E and the associations E--$>$ P(E) get 
characterized and created in the first place. There is an infinite continuum
of projection operators P such that S--$>$ PSP would generate a new state.
Why are some particular P's given favored status, and why are these favored
P's associated with ``experiences''?

This favored status is this: some one of these favored  P's will 
be picked out from the continuum of possibilities, in conjunction with the 
next phase of the dynamical process. This next phase is the putting to 
Nature of the question: Does the current state S  jump to PSP or not? 

To provide some basis for getting the universe going in a way that
tends to produce stable or enduring structure, instead of mere chaotic 
random activity, I assume that a basic characteristic of the underlying 
dynamics is to select only projectors P that impose a certain repetitiveness
on the dynamical structure. These qualities of repetitiveness are assumed 
to be fundamental qualities of the projectors. But each such quality is a 
characteristic that is more general in its nature than any 
particular realization of it. These general qualities I call ``feels'': 
they encompass all human experiences, but extend far beyond.  

Thus the basic assumption is that certain projectors P have ``feels'', but most
do not, where a ``feel'' is a generalized version of a human experience. Each
feel is characterized by a quality of repetitiveness, and the actualization
of this feel entails the actualization of some particular realization of that
quality or pattern of repetitiveness within the dynamical structure that 
constitutes the universe. This actualization is expressed by the 
transformation S--$>$ PSP where P = P(E), and E is the feel: it is the 
quality of the repetitiveness that is being actualized. 

This general tendency to produce repetitive spatio-temporal patterns 
carries over to human experience, and will, I believe, be greatly enhanced 
by natural selection within the biological sphere. Thus the selection, from 
among the proferred potential experiences, of the next $E'$, will be such as 
to favor a sequences E--$>$ P(E)--$>$ $E'$ such that $E'$ is either the same 
as E, or at least the same as E in some essential way. Thus experiences, and 
their more general ontological cousins, feels, are tied to the generation of 
self-reproducing structures. This generation of regenerating/reverberating 
stable structures underlies quantum dynamics, in the form of the creation by 
the dynamics of stable and quasi-stable particles, and extends beyond human 
beings, both to biological systems in general, and even to the overall 
organization of the universe, according to the ideas being developed here.

As regards this repetitiveness, it is undoubtedly pertinent that classical 
mechanics is formulated basically in space-time, with lawfulness expressed 
essentially by a static or quasi-static quality of momentum-energy. But 
the essence of the transition to quantum theory is precisely that this 
static quality of momentum-energy is replaced by a repetitive quality, by 
a characteristic oscillatory behavior': quantum theory is basically about 
repetitive regeneration.

In line with all this, I assume that the projection operators P act by 
specifying  the (expectation values of the) quantum elecromagetic field. 
There are many reason for believing that this is the way nature operates:

1. The EM fields naturally integrate the effects of the motions of the 
billions of ions and electrons that are responsible for our neural processes.
Thus examining the EM fields provide a natural way of examining the state of 
the brain, and selecting a state of the EM field of the brain provides a 
natural way of controlling the behavior of the brain.

2. The EM field has marvelous properties as regards connections to classical
physics. The bulk of the low-energy EM state automatically organizes 
itself into a superposition of ``coherent states'', each of which is 
described by a classical electromagnetic field, and which enjoys many 
properties of this classical elecromagnetic field. These ``classical'' 
states are brought into the dynamical structure in a natural way: the 
condition that each actually realized state will correspond to essentially 
a single one of these classically describable coherent states is what is 
needed to deal effectively, in a physically realistic way, with the 
infra-red divergence problem in quantum electro-dynamics. 
[See Stapp (1983), and Kawai and Stapp (1995)]

3, These ``classical'' states (coherent states) of the quantum EM field 
are robust (not easily disrupted by the thermal and random noises in a 
warm wet brain): they are ideal for use in generating self-reproducing 
effects in a warm, wet, noisy enviroment. [See Stapp (1987), (1993, p.130), 
and Zurek (1993)]
 
4. These classical states are described by giving the ampitudes in each
of the oscillatory modes of the field: spacetime structure arises from
phase relationships among the different oscillatory modes.

Although the theory being developed here maintains a close connection
to classical physics, its logical and ontological structure is very 
different. In classical physics the dynamics is governed entirely by 
myopic local rules: i.e., by rules that specify the evolution of everything 
in the universe by making each local variable respond only to the physical 
variables in its immediate neighborhood. Human experiences are thus 
epiphenomenal in the sense that they do not need to be recognized as 
entities that play any dynamical role: the local microscopic description, 
and the local laws, are sufficient to specify completely the evolution of 
the state of physical universe. Experiential gestalts can regarded as
mere effects of local dynamical causes, not as essential elements in the
causal progession.

But the most profound lesson about nature learned in the twentieth century 
is that the empirically revealed structure of natural phenomena cannot be 
comprehended in terms of any local dynamics: natural phenomena are strictly 
incompatible with the idea that the underlying dynamics is local. 

The second most profound lesson is that the known observed regularities
of natural phenomena can be comprehended in terms of a mathematical
model built on a structure that behaves like representations of knowledge, 
rather than representations of matter of the kind postulated to exist in 
classical mechanics: the carrier of the structure that accounts for the 
regularities in nature that were formerly explained by classical physical
theory is, according to contempory theory, more idealike than matterlike, 
although it does exhibit a precise mathematical structure. 

The third essential lesson is that this new description, although complete
in important practical or pragmatic ways, is, as an ontological description,
incomplete: there is room for additional specifications, and indeed {\it an
absolute need for additional specifications} if answers are to be given to 
questions about how our experiences come to be what they are. The presently
known rules simply do not fix this aspect of the dynamics. The purpose of 
work is to make a first stab at filling this lacuna.

One key point, here, is that brains are so highly interconnected that it will
generally be only large macroscopic structures that have a good chance of
initiating a causal sequence that will be self-reproductive. So each 
possible experience E should correspond to a P(E) that creates a macroscopic 
repetitiveness in the states of a brain.

A second key point is that our knowings/experiences can be efficacious not 
only in the sense that they select, in each individual case, what actually 
happens in that case, but also in the statistical sense that the rules that
determine which questions are put to Nature, can skew the statistical 
properties, even if the answers to the posed questions follow the quantum 
statistical rules exactly. I turn now to a discussion of this point and its 
important consequences.

\vskip .1in
\noindent {\bf 12. Measurements, Observations, and Experiences.}
\vskip .1in

A  key  question  is  whether, in  a  warm wet  brain,  collapses
associated with knowings would have any effects that are different from what
would be predicted by classical theory, or more precisely, by a Bohm-type 
theory. Bohm's theory yields all the predictions of quantum theory in a way
that, like classical mechanics, makes consciousness epiphenomenal: 
the flow of consciousness is governed deterministically (but nonlocally) 
by a state of the universe that evolves, without regard to consciousness, 
in accordance with local deterministic equations of motion. Bohm's theory, 
like classical physics, tacitly assumes a connection between consciousness 
and brain activity, but the details of this connection are not specified.

The aim of the present work is to specify this connection, starting from 
the premise that the quantum state of the universe is essentially a 
compendium of knowledge, of some general sort, which includes all human 
knowledge, as contrasted to something that is basically mechanical, and 
independent of human knowledge, like the quantum state in Bohmian 
mechanics.  

I distinguish a ``Heisenberg collapse'', S--$>$ PSP or 
S--$>$ (1-P)S(1-P), from a ``von Neumann collapse'' S--$>$[PSP + (1-P)S(1-P)].
 The latter can be 
regarded as either a precursor to the former, or a representation of 
the {\it statistical} effect of the collapse: i.e., the effect if one 
averages, with the appropriate weighting, over the possible outcomes.

This latter sort of averaging would be pertinent if one wanted to examine
the observable consequences of assuming that a certain
physical system is, or alternatively is not, the locus of collapses.

This issue is a key question:  Are there possible empirical distinctions
between the behaviors of systems that are---or alternatively are not--- 
controlled by high-level collapses of the kind that this theory associates 
with consciousness. Can one empirically 
distinguish, on the basis of theoretical principles, whether collapses
of this kind are occurring within some system that is purported to be
conscious. This question is pertinent both to the issue of whether some 
computer that we have built could, according to this theory, be conscious, 
and also to the issue of whether our own behavior, as viewed from the outside,
has aspects that reveal the presence of the sort of quantum collapses that this
theory associates with consciousness.

This question about differences in behaviour at the statistical level feeds
also into the issue of whether being conscious has survival value. If 
behaviour has, on the average, no  dependence on whether or not
collapses occur in the system then the naturalistic idea that consciousness 
develops within biological systems due to the enhancement of survival
rates that the associated collapses provide would become nonsense. 
Indeed, that idea is nonsense within classical physics, for 
exactly this reason: whether conscious thoughts occur in association with 
certain physical activities  makes absolutely no difference 
to the microlocally determined physical behavior of the system.

There are certain cases in which a von Neumann collapse, 
S--$>$ [PSP + (1-P)S(1-P)], would produce no observable effects on 
subsequent behavior. To understand these conditions let us examine the 
process of measurement/observation. 

If one separates the degrees of freedom of the universe into those of 
``the system being measured/observed'', and those of the rest of the 
universe, and writes the state of the universe as 
$$
 S = |\Psi >< \Psi | 
$$
with
$$ 
|\Psi> = \sum_i \phi_i \chi_i,
$$
where the $\phi_i$ are states of ``the system being measured/observed'', 
and the $\chi_i$ are states of the rest of the universe, then since we 
observers are parts of the rest of the universe it is reasonable to demand 
that if someone can have an experience E then there should be a basis of 
orthonormal states $\chi_i$ such that the corresponding projector P(E) is 
defined by
$$
P(E) \phi_i = \phi_i
$$
for all $i$,
$$
P(E) \chi_i = \chi_i 
$$
for $i$ in $I(E)$, but
$$
P(E) \chi_i = 0,
$$ 
otherwise, where $I(E)$ is the set of indices $i$ that label those states 
$\chi_i$ that are compatible with experience E. 

A ``good measurement'' is
defined to be an interaction between the system being measured and the
rest of the universe such that the set of states $\phi_i$ defined above
with $i$ in $I(E)$ span a proper subspace of the space corresponding
to the measured system. In this case the knowledge that 
$i$ is in the set $I(E)$ ensures that the state of the measured system 
lies in the subspace spanned by the set of states $\phi_i$ with $i$ in 
$I(E)$. That is, experience E would provide knowledge about the 
measured system. 

Let P\_ be the projector that projects onto the subspace spanned
by the set of states $\phi_i$ with $i$ in $I(E)$. Then a von Neumann 
collapse with P\_ in place of P would be identical to the von Neumann 
collapse S--$>$ [PSP + (1-P)S(1-P)]. But then the observer would be unable to
determine whether a collapse associated with P\_ occurred in the system,
unbeknownst to him, or whether, on the contrary, the definiteness of the
observed outcome was brought about by the collapse associated with his
own experience. This is essentially von Neumann's conclusion. 

But why should an actual collapse associated with the measured/observed system
correspond in this special way to a subsequent experience of some human 
being? Why should an actually occurring P\_ be such as to ensure an 
equivalence between P\_ and a P(E)?

Von Neumann's approach to the measurement problem suggests that such a 
connection would exist.

In both the von Neumann and Copenhagen approaches the measuring device
plays a central role. Different perceptually distinguishable locations
of some ``pointer'' on the device are supposed to become correlated, during 
an interaction between the measured system and the measuring device, to 
different orthogonal subspaces of the Hilbert space of the measured system.
This perceptual distinctness of the possible pointer positions means that
there is a correlation between pointer {\it locations} and experiences. That
connection must be explained by the theory of consciousness, which is
what is being developed here. But why, ontologically, as opposed to
epistemologically, should the projector P\_ in the space of the 
measured/observed system be to a state that is tied in this way to
something outside self, namely the {\it location} of a pointer on a measuring 
device with which it might have briefly interacted at some earlier time. 

Von Neumann did not try to answer this question ontologically.
If the real collapse were in the brain, and it corresponded to seeing the
pointer at some one of the distinguishable locations, then from an 
epistemological point of view the effect of this collapse would be equivalent
to applying P\_ to the state of the measured/observed system.

If one works out from experiences and brains, in this way. one can formulate
the collapses in terms of collapses out in the world, instead of inside
the brain, and largely circumvent (rather than resolve) the mind-brain 
problem. Then the equivalence of the experience to the collapse at the 
level of the measured/observed system would become true essentially by 
construction: one defines the projectors at the level of the
measured/observed system in a way such that they correspond to the distinct
perceptual possibilities. 

But from a non-subjectivist viewpoint, one would like to have a 
characterization of the conditions for the external collapse that do 
not refer in any way to the observers.

One way to circumvent the observers is to use the fact that the pointer
interacts not only with observers but also with ``the environment'', which 
is imagined to be described by degrees freedom that will never be measured 
or observed.  The representation of S given above will again hold with 
the $\phi_i$ now representing the states of the system being measured plus 
the measuring device, and the $\chi_i$ corresponding to states of the 
environment.

The interaction between the pointer and the environment should quickly
cause all the $\chi_i$ that correspond to different distinct locations of 
the pointer to become orthogonal. 

All observable projectors  P are supposed to act nontrivially only on the 
states $\phi_i$: they leave unchanged all of the environmental states $\chi_i$.But then all observable aspects of the state S reside in tr S, where tr
stands for the trace over the environmental degrees of freedom.

Let $P_i$ be a projector onto an eigenstate of tr S. Suppose one postulates 
that each of the allowed projectors P\_ is a sum over some subset of the 
$P_i$, or, equivalently that each possible P\_ commutes with tr S, and is
unity in the space of the degrees of freedom of the environment.

This rule makes each allowed P project onto a statistical mixture of pointer
locations, in cases where these locations are distinct. So it give the sort of
P's that would correspond to what observers can observe, without mentioning 
observers. 

The P's defined in this way commute with S. But then the effect of any
von Neumann reduction is to transform S into S: the von Neumann reduction
has no effect at all. The collapse would have no effect at all on the 
average over the alternative possible answers to the question of whether or
not the collapse occurs. This nondependence of the average is of course an
automatic feature of classical statistical mechanics.

The theory being described here is a development of von Neumann's
approach in the sense that it gives more ontological reality to the 
quantum state than the Copenhagen approach, and also in the sense that it
follows von Neumann's suggestion (or what Wigner describes as von 
Neumann's suggestion) of bringing consciousness into the theory as a 
real player. But it differs from the models discussed above that are based 
on his theory of measurement. For it does not associate collapses with
things like positions of pointers on measuring devices. The projectors
P(E) associated experiences E are in terms of classical aspects of
the electromagnetic fields in brains of observers. That would be in 
line with von Neumann's general idea, but he did not go into details 
about which aspects of the brain were the pertinent ones. Rather he
circumvented the issue of the mind-brain connection by centering his
attention on the external devices and their pointer-type variables. 

The classical aspects of the EM field are technically different from 
pointers because their interaction with the environment is mainly their 
interaction with the ions and electrons of the brain, and these are
the very interactions that both create these aspects of these fields,
and that are in part responsible for the causal effects of the 
experiences E through the action of the projectors P(E). So what was
formerly an uncontrolled and unobservable environment that disturbed
the causal connections is now the very thing that creates the coherent
oscillatory structure through which our experiences control our brains.

The effects of this switch will be examined in the next section.

\vskip .1in
\noindent   {\bf 13. Efficacy of Knowings.}
\vskip .1in

A formalism for dealing with the classical part of the the 
electro-magnetic field, within quantum electrodynamics (QED), has been
developed in Stapp (1983) and Kawai and Stapp (1995), where it was shown
that this part dominates low-energy aspects, and is exactly expressed 
in terms of a unitary operator that contains in a finite way the terms 
that, if not treated with sufficient precision, lead to the famous infrared
divergence problem in QED. This classical part is a special kind of
quantum state that has been studied extensively. It is a so-called
coherent state of the photon field. Essentially all of the low-energy
contributions are contained within it, and the effects of emission and 
re-absorption are all included. However, different classically conceived
current sources produce different ``classical fields'', and hence the
full low-energy field is a quantum superposition of these classical states.

Each such classical state is a combination (a product) of components each 
of which has a definite frequency. All of the electrons and ions in the
brain contribute to each of these fixed frequency components, with
an appropriate weighting determined by that frequency. Thus the 
description is naturally in the frequency domain, rather than in spacetime
directly: spatial information is encoded in quantum phases of the various
fixed frequency components. Each value is represented, actually, by
a gaussian wave packet centered at that value, in a certain space, and hence
neighboring values are represented by overlapping gaussian wave packets.

To exhibit a basic feature I consider a system of just three of these states.
Suppose state 2 has all of the correct timings to elicit some coordinated 
actions. It represents in this simple model the state singled out by
the projector P = P(E). Suppose it is dynamically linked to some motor
state, represented by state 3: the dynamical evolution carries 2 to 3.
Let state 1 be a neighbor of state 2 such that the dynamical evolution
mixes 1 and 2. (I use here the Schroedinger picture, for convenience.)

The transition from 2 to 3 will tend to depopulate the coupled pair
1 and 2. This depopulation of the system 1 and 2 will occur naturally
whether or not any von Neumann collapse associated with P occurs.
The question is: Can a von Neumann collapse associated with P
affect in a systematic way the rate of depopulation from the coupled
pair 1 and 2. The answer is ``Yes'': it can speed up the emptying of the
amplitude in the system 1 and 2 into the system 3 that represents the
motor action. This means that the effect of repeatedly putting to nature 
the question associated with P can have the effect of producing the motor 
action more quickly than what the dynamics would do if no question was put:
putting the question repeatedly can effect the probabilities, compared Bohm's 
model, in which there are no collapses. The quantum rules regarding the 
probability of receiving a `Yes', or alternatively a `No', are stricly 
observed.

To implement the dynamical conditions suppose the initial state is 
represented, in the basis consisting of our three states 1, 2, and 3,
by the Hermitian matrix S with $S_{1,1} = x$, $S_{2,2} = y$,
$S_{1,2}= z$, $S_{2,1}=z^*$, and all other elements zero. Suppose
the coupling between states 2 and 3 is represented by the unitary
matrix U  with elements $U_{1.1} = 1$, and 
$$
U_{2,2}=U_{2,3}=U_{3,3}= -U_{3,2}= r = (2)^{-1/2},
$$
with all other elements zero.

The mixing between the states 1 and 2 is represented by the unitary
matrix M with 
$$
M_{1,1}=c, M_{1,2}= s, M_{2,1}= -s^*, M_{2,2}=c^*, M_{3,3}=1,
$$
with all other elements zero. Here $c^* c + s^* s$ = 1.

The initial probability to be in the state 2 is given 
Trace PS = y, where P projects onto state 2.
The action of U depopulates state 2: $Trace PUSU^{-1}= y/2$.

Then the action of the mixing of 1 and 2 generated by M brings
the probability of state 2 to 
$$
Trace PMUSU^{-1}M^{-1} = (xs^* s)+ (yc^* c/2) -zcs^*r -z^*c^*sr,
$$
where r is one divided by the square root of 2

For the case c = s = r this gives for the probability of state
2:
$$ 
(xs^* s) + (yc^* c/2) -zcs^*r - z^*c^*sr = x/2 + y/4 - zr/2 -z^*r/2
$$

Since states 1 and 2 are supposed to be neighbors the most natural
initial condition would be that the feeding into these two states
would be nearly the same: the initial state would be a super position
of the two states with almost equal amplitudes. This would make x = y = 
z = $z^*$.
Then the probability of state 2 becomes
$$
prob = y/2 + y/4 -yr
$$
Then the effect of the mixing M is to decrease from y/2 the 
probability in the state 2 that feeds the motor action. 

If the question E, with P(E)= P, is put to nature before U acts, then the
effect of the corresponding von Neumann reduction is to set z to zero.
Hence in this case
$$
prob = y/2 + y/4,
$$
and the probability is now increased from y/2. 

Thus putting the question to Nature speeds up the motor response, on 
the average, relative to what that speed would be if the question were not
asked.

The point of this calculation is to establish that this theory allows 
experiences to exercise real control over brain activity, not only by 
making the individual choices between possibilities whose probabilities 
are fixed by the quantum rules, but also at a deeper level by shaping, 
through the choices of which questions are put to nature, those 
statistical probabilities themselves. This opens the door both to
possible empirical tests of the presence of collapses of the kind 
predicated in this theory, and to a natural-selection-driven co-evolution 
of brains and their associated minds. 

\vskip .1in
\noindent   {\bf 14. Natural Selection and the Evolution of Consciousness.}
\vskip .1in

In a naturalistic theory  one would not expect consciousness to be
present in  association with a  biological system unless  it had a function:
nothing as complex and refined as  consciousness should be present unless it
enhances the survival prospects of the system in some way. 

This requirement poses a problem for  a classically described system because
there  consciousness is causally  non-efficatious: it is  epiphenomenal. Its
existence  is not,  under  any  boundary  conditions,  {\it  implied} by the
principles of classical  physics in the way that  what we call ``a tornado''
is , under  appropriate  boundary  conditions, implied by  the principles of
classical physics.  Consciousness  could therefore be  stripped away without
affecting  the behavior  of the  system in any  way. Hence it  could have no
survival value.

Consider two  species,  generally on a par,  but such that  in the first the
survival-enhancing  templates for action are  linked to knowings, in the way
described  above, but in  the second  there is no  such  linkage. Due to the
enhancement  effects described in the  preceding section  the members of the
first species will  actualize their  survival-enhancing templates for action
faster and more often  than the members of the  second species, and hence be
more likely to survive. And over the course of generations one would expect
the organism to evolve in such a way that the possible experiences E 
associated with it, and their consequences specified by the associated 
projection operators P(E), will become ever better suited to the survival 
needs of the organism.

\vskip .1in
\noindent   {\bf 15. What is Consciousness?}
\vskip .1in

When  scientists who study  consciousness are asked to define what
it is they study, they are reduced either to defining it in other words that
mean  the same   thing, or to   defining it   ostensively  by  directing the
listener's  attention to  what the word stands  for in his own life. In some
sense   that is  all one  can do  for any  word:  our  language is  a web of
connections  between our experiences of various kinds, including sensations,
ideas,  thoughts, and theories.

If we were to ask a  physicist of the last  century what an ``electron'' is,
he could  tell us about  its  ``charge'', and its  ``mass'',  and maybe some
things  about its  ``size'', and how  it is  related to  ``atoms''. But this
could  all be some  crazy  abstract   theoretical idea,  unless a  tie-in to
experiences is  made. However, he  could give a lengthy  description of this
connection, as  it was  spelled out by  classical physical  theory. Thus the
reason that a  rational  physicist or  philosopher of the  ninteenth century
could believe that  ``electrons'' were real, and  perhaps even ``more real''
than our thoughts about them, is that they were understandable as parts of a
well-defined  mathematical framework  that  accounted---perhaps not directly
for our  experiences themselves, but  at least---for how  the contents of 
our experiences hang together in the way they do.

Now, however, in the  debate between materialists  and idealists, the tables
are  turned: the  concepts of  classical  physics,  including  the classical
conception  of tiny  electrons  responding  only to  aspects of  their local
environment, absolutely cannot account for the macroscopic phenomena that we
see  before our  eyes. On  the  contrary: the  only  known theory  that does
account for all the empirical phenomena, and that is  not   burdened with
extravagent   needless  ontological  excesses,  is a  theory that  is neatly
formulated directly in terms of our knowings. So the former reason for being
satisfied   with the  idea of  an  electron,  namely  that  it is  part of a
parsimonious  mathematical  framework that  accounts  quantitatively for the
contents of our  experiences, and gives us a  mathematical representation of
what persists  during the   intervals  between our   experiences, has
dissolved  insofar as it  applies to the  classical idea of  an electron: it
applies now, instead, to our knowings, and the stored compendium of all 
knowings, the Hilbert space state of the universe.

To elicit   intuitions, the  classical  physicist  might have  resorted to a
demonstration of tiny ``pith balls'' that attract or repel each other due to
(unseen) electric fields, and then  asked the viewer to imagine much smaller
versions  of what he  sees  before his  eyes. This  would give  the viewer a
direct intuitive basis for thinking he understood what an electron is.

This intuitive  reason for the  viewer's being  satisfied with the notion of
an electron  as an  element of  reality is that  it was a  generalization of
something very  familiar: a  generalization of the tiny  grains of sand that
are so common in our ordinary experience, or of the tiny pith balls.

No things are  more familiar  to us than our  own  experiences. Yet they are
elusive: each of  them disappears  almost as soon as it  appears, and leaves
behind only a fading impression, and fallible memories. 
 
However, I shall try in this section to nail down a more solid idea  of what
a conscious experience is: it unifies the theoretical  and intuitive aspects
described above.
 
The metaphor is the experienced sound of a musical chord.

We have  all  experienced how a  periodic beat  will, when the  frequency is
increased, first be heard as a closely spaced sequence of individual pulses,
then as a buzz, then as  a low tone, and then as  tones of higher and higher
pitch.  A tone  of  high  pitch, say  a high  C, is not experienced by most
listeners as a sequence of finely  spaced individual pulses,  but as
something experientially unigue.

The same goes for major and minor  chords: they are experienced differently,
as a different  gestalts. Each chord,  as normally  experienced, has its own
unique total quality,  although an experienced  listener can attend to it in
a way that may reveal the component elements. 

One can generalize  still further to the complex  experience of a moment  of
sound in a Beethoven symphony. 

These  examples show  that a  state that  can be  described  physically as a
particular combination  of vibratory motions is  experienced as a particular
experiential quality: what we cannot  follow in time, due to the rapidity of
the variations, is experienced as a gestalt-type impression that is a quality
of the entire distribution of  energy among the sensed frequencies.

According  to the theory purposed here, the aspect  of brain  dynamics that
corresponds to a conscious experience  is a complex pattern of reverberating
patterns of EM  excitations that  has reached a  stable steady state and
become a template for immediate further brain action. Its actualization by a
quantum event  initiates that action: it selects out of an infinite of
alternative   competing and  conflicting  patterns of  neural  excitations a
single   coherent  energetic  combination  of   reverberating  patterns that
initiates, quides, and monitors, an  ongoing coordinated evolution of neural
activities.   The  experience  that  accompanies  this   suddenly-picked-out
``chord'' of reverberations is, I  suggest, the ``quality'' of this complex 
pattern of  reverberations. Because the sensed combinations of EM 
reverberations that constitute the template for action is far more complex
than those that represent auditory sounds, the quality of the former chord 
must be far more complex than that of the latter.

But the most important quality of our experiencess is that they have 
meanings. These meanings arise from their intentionalities, which 
encompass both intentions and attentions. The latter are intentions to attend 
to---and thereby to update the brains representation of---what is attended to.

These aspects of the experience arise from their self-reproducing
quality: their quality of re-creating themselves. In the case
of our human thoughts this self-reproductive feature has evolved to the
point such that the present thought contains a representation of what
will be part of a subsequent thought: the present experience E contains
an image of a certain prolongation (projection into the future) of the current 
Jamesian sequence of temporal components that is likely, by virtue of 
the causal effect of E, namely S--$>$ PSP, with P = P(E), to be the 
current Jamesian sequence of a subsequent experience $E'$. 

Thus the meaning of the experience, through physically imbedded in the present
state of the brain that it engenders, consists of the image of the future 
that it is likely to generate, within the context of its fringe. 

\vskip .2in
{\bf Acknowledgements}

This article is essentially a reply to detailed questions about
earlier works of mine raised by Aaron Sloman, Pat Hayes, Stan Klein, 
David Chalmers, William Robinson, and Peter Mutnick. I thank them for 
communicating to me their dissatisfactions. I also thank Gregg Rosenberg 
and John Range for general support.
\vskip .2in
{\bf References.}

Bohr, N (1934), {\it Atomic Theory and the Description of Nature}
              (Cambridge: Cambridge University Press).

Bunge, M. (1967), {\it Quantum Theory and Reality} (Berlin: Springer). 

Einstein, A. (1951) {\it Albert Einstein: Philosopher-Scientist}
             ed. P.A. Schilpp (New York: Tudor).

Fogelson, A.L. \& Zucker, R.S. (1985),`Presynaptic calcium diffusion
from various arrays of single channels: Implications for transmitter
release and synaptic facilitation', {\it Biophys. J.}, {\bf 48}, 
pp. 1003-1017.

Feynman, R., Leighton, R., and Sands, M., (1965) {\it The Feynman
Lectures in Physics.} (Vol. III, Chapter 21).(New York: Addison-Wesley).

Haag, R. (1996) {\it Local Quantum Physics} (Berlin: Springer).p 321.

Heisenberg, W. (1958a) `The representation of nature in contemporary 
    physics',  {\it Deadalus} {bf 87}, 95-108.

Heisenberg, W. (1958b) {\it Physics and Philosophy} (New York: Harper and 
     Row).

Hendry, J. (1984) {\it The Creation of Quantum Theory and the Bohr-Pauli
      Dialogue} (Dordrecht: Reidel).

Kawai, T and Stapp, H.P. (1995) `Quantum Electrodynamics at large distance I, 
II, III', {\it Physical Review}, {\bf D 52} 3484-2532.

Joos, E. (1986) `Quantum Theory and the Appearance of a Classical World',
    {\it Annals of the New York Academy of Science} {\bf 480} 6-13.

Omnes, R. (1994) {\it The Interpretation of Quantum Theory}, (Princeton: 
Princeton U.P.) p. 498.

Stapp, H.P. (1975) `Bell's Theorm and World Process', {\it Nuovo Cimento}
{\bf 29}, 270-276. 
 
Stapp, H.P. (1977) `Theory of Reality',  {\it Foundations
of Physics} {\bf 7}, 313-323.

Stapp, H.P. (1979) `Whiteheadian Approach to Quantum Theory', {\it Foundations
of Physics} {\bf 9}, 1-25.

Stapp, H.P. (1983) `Exact solution of the infrared problem' {\it Physical
Review}, {\bf 28} 1386-1418.

Stapp, H.P. (1993) {\it Mind, Matter, and Quantum Mechanics}
(Berlin: Springer), Chapter 6.\\
\& http://www-physics.lbl.gov/\~{}stapp/stappfiles.html

Stapp, H.P. (1996) `The Hard Problem: A Quantum Approach', {\it Journal
of Consciousness Studies}, {\bf 3} 194-210.

Stapp, H.P. (1997) `Nonlocal character of quantum theory', {\it American
Journal of Physics}, {\bf 65}, 300-304.\\
For commentaries on this paper see:\\
http://www-physics.lbl.gov/\~{}stapp/stappfiles.html\\ 
The papers quant-ph/9905053 cited there can be accessed at\\
quant-ph@xxx.lanl.gov by putting in the subject field the command:\\
get 9905053

Stapp, H.P. (1997a) `Science of Consciousness and the Hard Problem',
{\it J. of Mind and Brain}, vol 18, spring and summer.

Stapp, H.P. (1997b) `The Evolution of Consciousness',\\
http://www-physics.lbl.gov/\~{}stapp/stappfiles.html

Wigner, E. (1961) `The probability of the existence of a self-reproducing
      unit', in {\it The Logic of Personal Knowledge} ed. M. Polyani
     (London: Routledge \& Paul) pp. 231-238. 

Zucker, R.S. \& Fogelson, A.L. (1986), `Relationship between transmitter 
release and presynaptic calcium influx when calcium enters through disrete
channels', {\it Proc. Nat. Acad. Sci. USA}, {\bf 83}, 3032-3036.

Zurek, W.H. (1986) `Reduction of the Wave Packet and Environment-Induced
Superselection', {\it Annals of the New York Academy of Science} {\bf 480}, 
89-97

Zurek, W.H., S. Habib, J.P. Paz, (1993) `Coherent States via Decoherence',
Phys. Rev. Lett. {\bf 70} 1187-90.

\newpage
\noindent {\bf Appendix A. Quantum Effect of Presynaptic Calcium 
             Ion Diffusion.}

Let me  assume here, in  order to  focus  attention on a  particular  easily
analyzable source  of an important  quantum effect, that  the propagation of
the action potential along nerve fibers is well represented by the classical
Hodgson-Huxley   equation, and  that  indeed all of  brain  dynamics is well
represented by the classical approximation apart from one aspect, namely the
motions of the pre-synaptic calcium ions from the exit of the micro-channels
(through which they have entered the  nerve terminal) to their target sites.
The   capture  of  the ion  at  the   target  site   releases a   vesicle of
neurotransmitter into the synaptic cleft.

The  purpose of the  brain activity  is to  process clues  about the outside
world coming  from the  sensors, within the  context of a  current  internal
state representing the individual's state of readiness, in order  to produce
an appropriate  ``template for  action'', which can then  direct the ensuing
action  (Stapp, 1993).  Let it be  supposed that  the  classically described
evolution  of the  brain,  governed by the  complex  nonlinear  equations of
neurodynamics,  will  cause the brain  state move  into the  vicinity of one
member  of a set of attractors. The various attractors represent the various
possible   templates for action:  starting from this  vicinity, the state of
the  classically   described  body/brain will  evolve through  a sequence of
states that  represent  the  macroscopic course of  action specified by that
template for action.

Within  this   classically   described  setting  there are  nerve  terminals
containing the  presynaptic calcium ions. The  centers of mass of these ions
must be   treated as quantum  mechanical  variables. To  first approximation
this means  that each of these  individual calcium ions is represented as if
it were a  statistical ensemble of  classically conceived calcium ions: each
individual   (quantum)  calcium ion  is  represented as a  cloud or swarm of
virtual classical calcium ions all existing together, superposed. This cloud
of  superposed   virtual copies  is called  the wave  packet.  Our immediate
interest is in the  motion of this  wave packet as it moves from the exit of
a microchannel  of  diameter  1 nanometer to  a target  trigger site for the
release of a vesicle of  neurotransmitter into the synaptic cleft. 

The  irreducible  Heisenberg  uncertainty in the  velocity of  the ion as it
exits the   microchannel is  about $1.5$  m/sec,  which is  smaller than its
thermal velocity  by a factor of  about $4 \times  10^{-3}$. The distance to
the      target      trigger    site   is    about    $50$       nanometers.
(Fogelson,1985;Zucker,1986) Hence the spreading of the wave packet is of the
order of  $0.2$  nanometers, which is  of the  order of the  size of the ion
itself, and of the target trigger  site. Thus the decision as to whether the
vesicle is  released or not,  in an  individual instance,  will have a large
uncertainty due to the large  Heisenberg quantum uncertainty in the position
of the calcium ion relative to the trigger site: the ion may hit the trigger
site and release the vesicle, or it may miss it the trigger site and fail to
release the vesicle. These two  possibilities, yes or no, for the release of
this vesicle by this ion  continue to exist, in a  superposed state, until a
``reduction of the wave packet'' occurs. 

If there is a situation in which a certain particular {\it set of  vesicles}
is released, due to the  relevant calcium ions  having been  captured at the
appropriate    sites,  then  there   will be  other   nearby  parts  of  the
(multi-particle)  wave  function of  the brain in  which some  or all of the
relevant captures do not take place---simply because, for those nearby parts
of the wave  function, the pertinent  calcium ions miss  their targets---and
hence the corresponding vesicles are not released. 

More generally, this means, in a situation that corresponds to a very  large
number N of  synaptic  firings, that, until  a reduction  occurs, all of the
$2^N$ possible  combinations of  firings and no firings  will be represented
with comparable  statistical weight  in the wave  function of the brain/body
and its environment. Different combinations of these firings and  no firings
can lead to different  attractors, and thence to  very different macroscopic
behaviours of the body that is being controlled by this brain.

The important  thing, here, is that  there is, {\it on  top of the nonlinear
classically described neurodynamics}, a quantum mechanical {\it  statistical
effect} arising from the spreading out of the wave functions  of the centers
of mass of the  various presynaptic  calcium ions  relative to  their target
trigger sites.The spreading out of  the wave packet is  unavoidable, because
it is a consequence of the Heisenberg  uncertainty principle. This spreading
is extremely important,  because it entails that  every vesicle release will
be   accompanied by a   superposed   alternative    situation of  comparable
statistical weight in  which that vesicle is not   released. This means that
wave  function of the  entire brain  must, as a  direct   consequence of the
Heisenberg  uncertainty  principle,  disperse  into  a shower  of superposed
possibilities   arising from  all the  different  possible   combinations of
vesicle  releases or   non-releases. Each   possibility can be   expected to
evolve into the neighborhood of some  one of the many different  attractors.
These different  attactors will be brain states  that will evolve,  in turn,
if no reduction  occurs, into  different possible  macroscopic  behaviors of
the brain and body.

Thus the effect  of the spreadings  of the wave  functions of the centers of
the presynaptic  calcium ions is  enormous: it will  cause the wave function
of the person's body in its environment to disperse, if no reduction occurs,
into a profusion of branches that represent all of the possible actions that
the  person is  at all  likely  to take  in the   circumstance at  hand. The
eventual    reduction of  the  wave  packet   becomes,  then,  the  decisive
controlling    factor:  in any  given  individual   situation the  reduction
selects---from    among  all  of the   possible   macroscopically  different
large-scale   bodily  actions   generated  by the   nonlinear  (and, we have
supposed, classically  describable)   neurodynamics---the single action that
actually occurs. 
  
In this discussion I have generated the superposed macroscopically different
possibilities by  considering only the spreading  out of the wave packets of
the  centers-of-mass of the pertinent  presynaptic  calcium ions relative to
the target trigger  sites, imagining the rest of  the brain neurodynamics to
be   adequately   approximated  by the  nonlinear   classically  describable
neurodynamics  of  the brain.  Improving upon this  approximation would tend
only to increase  the quantum effect I have described.

It  should be   emphasized  that  this  effect is   generated  simply by the
Heisenberg  uncertainty principle,  and hence cannot be  simply dismissed or
ignored  within a  rational  scientific  approach.  The effect  is in no way
dependent upon  macroscopic quantum coherence,  and is neither wiped out nor
diminished   by   thermal  noise.  The   shower of   different   macroscopic
possibilities  created by  this effect can  be reduced to  the single actual
macroscopic reality that we observe  only by a reduction of the wave packet.

\newpage 
{\bf Appendix B. Knowings, Knowledge, and Causality.}

I shall flesh out here the idea that Nature is built out 
of knowings, not matter.

A typical knowing of the kind that quantum theory is built upon
is a knowing that the pointer on the measuring device appears to lie
between the numbers 6 and 7 on the dial. This is the sort of fact
that all (or at least most) of science is built upon. It is quite
complex. The idea that the appearance pertains to a dial on something
that acts as a measuring device has a tremendous amount of education
and training built into it. Yet somehow this  knowing has this background
idea built into it: that idea is a part of the experience.

William James says about perceptions:

``Your acquaintance with reality grows literally by buds or drops 
of perception.  Intellectually and upon reflection you can divide 
these into components, but as immediately given they come totally 
or not at all.''

This fits perfectly with Copenhagen quantum theory, which takes these
gestalts as the basic elements of the theory. In the von Neumann/
Wigner type ontology adopted here there is, in association
with this knowing, a collapse of the state vector of the universe.
It is specified by acting on this state with a projection operator
that acts on the degrees of freedom associated with the brain of the
perceiver, and that reduces the state of the body/brain of the observer,
and consequently also the state of the whole universe, to the part of 
that state that is compatible with this knowing. 

So a knowing is a complex experiential type of event that, however, 
according to the theory, occurs in conjunction with a correspondingly 
complex ``physical'' event that reduces the state of the the brain/body 
of the person to whom the experience belongs to the part of that state 
that is compatible with the knowing. [I shall use the word ``physical''
to denote the aspect of nature that is represented in the Hilbert-space 
description used in quantum theory: this aspect is the quantum analog 
of the physical description of classical physics.]

That ``person'' is a system consisting of a sequence of 
knowings bound together by a set of tendencies that are specified by
the state of the universe. This state is essentially a compendium of 
prior knowings. However, these knowings are not merely human knowings, 
but more general events of which human knowings are a special case.

In strict Copenhagen interpretation quantum theory is regarded as merely a 
set of rules for making predictions about human knowledge on the basis of 
human knowledge: horses and pigs do not make theoretical calculations using
these ideas about operators in Hilbert space, and their ``knowings''
are not included in ``our knowldge. 

But in a science-based {\it ontology} it would be unreasonable to posit 
that human knowledge plays a singular role: human knowings must be 
assumed to be particular examples of a general kind of ``knowings'' that 
would include ``horse knowings'' and ``pig knowings''. These could be 
degraded in many ways compared to human knowings, and perhaps richer 
in some other dimensions, but they should still be of the same general 
ontological type. And there should have been some sort of things of this 
general ontological kind even before the emergence of life.
[In the section, ``What is Consciousness'', I have tried to provide an 
intuition about what a knowing associated with a nonbiological system
might be like.]

Science is an ongoing endeavor that is expected to develop ever more  
adequate (for human needs) ideas about the nature of ourselves and of 
the world in which we find ourselves. Newton himself seemed to 
understand this, although some of his successors did not. But the 
present stage of theoretical physics makes it clear that we certainly 
do not now know all the answers to even the most basic questions: physics 
is still very much in a groping stage when it comes to the details of 
the basic underlying structure. So it would be folly, from a scientific 
perspective, to say that we must give specific answers now to all
questions, in the way that classical physics once presumed to do. 

This lack of certainty is highlighted by the fact that the 
Copenhagen school could claim to give practical rules that worked in the
realm of human knowledge without paying any attention to the question
of how nonhuman knowings entered into nature. And no evidence contrary to
Copenhagen quantum theory has been established. This lack of data about
nonhuman knowledge would make it presumptuous, in a science-based approach,
to try to spell out at this time details of the nature of nonhuman knowings,
beyond the reasonable presumption that animals with bodies structurally 
similar to the bodies of human beings ought, to the extent they also
behave like human beings, to have similar experiences. But knowings cannot 
be assumed to be always exactly the kinds of experiences that we human beings 
have, and they could be quite different.

The knowings that I mentioned at the outset were percepts: knowings 
that appear to be knowings about things lying outside the person's body. 
But, according to the von Neummann/ Wigner interpretation, each such knowing 
is actually connected directly to the state of the person's 
body/brain, after that event has occurred. This state of the body/brain
will, in the case of percepts of the external world, normally be correlated 
to aspects of the state of the universe that are not part of the
body/brain. But experienced feelings, such as the feelings of warmth, joy,
depression, devotion, patriotism, mathematical understandings, etc. are not 
essentially different from percepts: all are experiences that are
associated with collapse events that reduce the state of the 
body/brain to the part of it that is compatible with the experience.. 

I have spoken here of a body/brain, and its connection to an experience. 
But what is this body/brain? It seems to be something different
from the knowing that it is connected to. And what is the nature of this
connection?

The body/brain is an aspect of the quantum mechanically described state 
of the universe. This Hilbert-space state (sometimes called density
matrix) is expressed as a complex-valued function of two vectors, each of 
which is defined over a product of spaces, each of which corresponds to 
a degree of freedom of the universe. Any system is characterized by a 
certain set of degrees of freedom, and the state of that system is 
defined by taking the trace of the state of the universe over all 
other degrees of freedom, thereby eliminating from this state any 
explicit reference to those other degrees of freedom. 

In this way the state of each system is separately definable, 
and dependent only on its own degrees of freedom, even though the system 
itself is basically only an aspect of the whole universe. Each part 
(i.e., system) is separately definable, yet basically ontologically 
inseparable from the whole: that is the inescapable basic message of 
quantum theory. Each system has a state that depends only on its own 
degrees of freedom, and this system, as specified by its state, is 
causally pertinent, because each knowing is associated with some system, 
and the probabilities for its alternative possible knowings are specified 
by its own state, in spite of the fact that the system itself is 
fundamentally an inseparable part of the entire universe. It is the
properties of the trace operation that reconciles these disparate 
requirements
 
The state of the universe specifies only the probabilities
for knowings to occur, and it generally undergoes an instantaneous
global instantaneous jump when a new knowing occurs. But this
probability, by virtue of the way it jumps when a new knowing occurs,
and suddenly changes in regions far away from the system associated 
with the new knowing, and that it is formulated in terms of infinite
sets of pssibilities that may never occur, is more like an idea or a 
thought than a material reality. Indeed, these properties of the state 
are exactly why the founders of quantum theory were led to the conclusion 
that the mathematical formalism that they created was about knowledge.

The state of the universe is the preserved compendium of all knowings.
More precisely, it is an aspect of that compendium that expresses certain
statistical properties pertaining to the next knowing. There is
presumeably some deeper structure, not captured by the properties
expressed in the Hilbert-space mathematical structure, that fixes what
actually happens. 

The knowings that constitute our experiences are the comings into being 
of  bits of knowledge, which join to form the knowledge that is 
represented by the state of the universe. This gives an ontology
based on knowings, with nothing resembling matter present. But the 
statistical causal structure of the sequence knowings is expressed in 
terms of equations that are analogs of the mathematical laws that 
governed the matter postulated to exist by the principles of classical 
mechanics. This connection to classical mechanics is enough to ensure a 
close similarity between the predictions of classical mechanics and those 
of quantum mechanics in many cases of interest, even though the two 
theories are based on very different mathematical structures.   

If one starts from the ontological framework suggested by classical
mechanics the questions naturally arise: Why should experiences exist
at all? And given that they do exist, Why should they be composed of
such qualities as sensations of (experiential) colors and (experiential)
sounds, and feelings of warmth and coldness, and perceptions of
simple geometric forms that correspond more directly to the shapes of
structures outside the body/brain than to  structures (such as patterns
of neural excitations that are presumably representing these various 
features) inside the body/brain. How do these experiential types of 
qualities arise in a world that is composed exclusively of tiny
material particle and waves? The experiential qualities are not 
constructible from their physical underpinnings in the way that  
all the physical properties of a tornado are, according to classical
mechanics, constructible from its physical constituents.  

Quantum theory allows one to get around these questions by eliminating
that entire classical ontology that did not seem to mesh naturally with
experiential realities, and replacing that classical ontology with one
built around experiential realities. These latter realities are embedded in 
a specified way, which is fixed by the pragmatic rules, into a mathematical 
structure that allows the theory to account for all the successes of 
classical mechanics without being burdened with its awkward ontological 
baggage.

A discussion of this appendix with cognitive scientist Pat Hayes can
be found on my website: \\ 
(http://www-physics.lbl.gov/`tilde'stapp/stappfiles.html), \\
where `tilde' stands for the tilde symbol.

\newpage
\noindent {\bf Appendix C. Quantum Wholism and Consciousness.}

One reason touted for the need to  use quantum theory in order to accomodate
consciousness  in  our  scientific  understanding of  brain  dynamics is the
seeming pertinence of quantum wholism  to the unitary or wholistic character
of the  conscious experience.

I shall here spell out that reason within the framework of a computer
simulation of brain dynamics.

Suppose  we consider  a field  theory of  the brain,  with  several kinds of
interacting fields, say, for example,  the electric and magnetic fields, and
a field   representing some  mass- and   charge-carrying field.  Suppose the
equations  of  motion  are local  and   deterministic.  This means  that the
evolution in time of each field value  at each spacetime point is completely
determined by the values of the various fields in the immediate neighborhood
of that spacetime point.  Suppose we can, with  good accuracy, simulate this
evolution with a huge collection of computers, one for each point of a cubic
lattice of finely spaced spatial  points, where each computer puts out a new
set of values for each the fields,  evaluated at that its own spatial point,
at each of a sequence of finely  spaced times. Each computer has inputs only
from the outputs of its  nearest few neighbors,  over a few earlier times in
the sequence of times. The outputs  are digital, and the equations of motion
are presumed to  reduce to  finite-difference equations  that can be readily
solved by the  stripped-down computers, which can  do only that. Thus, given
some  appropriate initial  conditions at  some early times,  this battery of
simple  digital  computers will  grind out  the  evolution of  the simulated
brain. 

Merely for definiteness I assume that the spatial lattice has a thousand
points along each edge, so the entire lattice has a billion points. Thus our
simulator has a billion simple computers.

Now suppose after some long time the field values should come to spell out a
gigantic  letter  ``M'':  i.e., the  fields  all vanish  except  on a set of
lattice points that have the shape of  a letter ``M'' on one of the faces of
the   lattice. If  the  outputs  are  printed  out at  the  location  of the
corresponding grid  point then you or  I, observing the  lattice, would know
that the letter  ``M'' had been formed. 

But would  the  battery of  dynamically  linked but   ontologically distinct
computers itself contain that  information explicitly? None of the computers
has  any   information in  its  memory  except   information  about  numbers
pertaining to its  immediate  neighborhood: each  computer ``knows'' nothing
except  what its  immediate  environment  is. So  nowhere in the  battery of
computers, B, has  the higher-level  information about  the global structure
been assessed and  recorded: the fact  that an ``M'' has  been formed is not
``known'' to the battery of  computers. Some other computer C, appropriately
constructed,  could examine  the outputs of  the various  elements of B, and
issue a correct statement about this global properties of B, but that global
information is not explicity expressed in the numbers that are recorded in B
itself: some extra processing would be needed for that. 

Of course, brains  examine  themselves. So B itself  might be able to do the
job that C did above, and issue the statement about its own global property,
and also record that information in  some way in the configuration of values
in the various simple computers: the  existence of this configuration can be
supposed  to have  been  caused by  the  presence of the  ``M'',  and can be
supposed to cause, under appropriate conditions, the battery of computers  B
to display on some lattice face the message: ``I did contain an `M' ".

So the information  about the global structure is  now properly contained in
the  structure of B, as  far as  causal  functioning is  concerned. But even
though the  configuration of values  that carries the  information about the
``M'' is correctly  linked causally  to past and future,  this configuration
itself {\it is} no  more than any  such configurations  was before, namely a
collection of tiny bits of information about tiny regions in space. There is
nothing in this  classical conception that  corresponds ontologically to the
entire  gestalt, ``M'', as a  whole. The  structure of  classical physics is
such that the  present reality is  specified by values  located within in an
infinitesimal interval centered on  the present instant, without any need to
refer to  any more  distant times. To  bring  relationships to  the past and
future events into the  present evolving  ontological reality would be alien
to the ideas  of classical  physics. There  is simply no  need to expand the
idea of reality in this way: it adds only superfluities to the ontologically
and dynamically closed structure of classical physics.
 
The situation  changes completely  when one quantizes  the system. To make a
computer  simulation of  the quantum  dynamics  one  generalizes the spatial
points  of the   classical  theory to   super-points. Each   possible entire
classical state is a  super-point. In our case,  each super-point is defined
by specifying at each of the points  in the lattice a possible value of each
of the several (in our  case three) fields. To  each super-point we assign a
super-computer. If  the number of  discrete allowed  values for our original
simple  computers was,  say, one  thousand  possible values  for each of the
three fields, and hence $10^9$ possible output values in all for each simple
computer, then the number of allowed classical states would be $10^9$ raised
to the power  $10^9$: each of  the $10^9$  simple computers  can have $10^9$
possible values. Thus  the number of needed  super-computers would be $10^9$
raised  to the  power  $10^9$.  In the  dynamical   evolution each  of these
super-computers   generates, in   succession, one  complex  number (two real
numbers) at each of the times in the finely spaced sequence of times. 

One can imagine that a  collapse event at some  time might make all of these
complex numbers, except one, equal to zero, and make the remaining one equal
to $1$. Then  the state would  be precisely  one of the  $10^9$ to the power
$10^9$  classical  states.  It would  then  evolve into  a  superposition of
possible classical  states until the next  collapse occurs. But the collapse
takes the state to a ``whole'' classical world. That is, each super-computer
is associated not just with some tiny region, but with the whole system, and
the collapses can be to states in which some whole region of spacetime has a
fixed    configuration of   values.  Thus, for   example,  there  would be a
super-computer  such  that its  output's being  unity would  mean that ``M''
appeared on one face. And the  collapse to that single state would actualize
that gestalt  ``M''. The sudden  selective creation of  this gestalt is more
similar  to  someone's  experiencing  this  gestalt than  any  occurrence or
happening in the classical dynamics,  because in both the experience and the
quantum  event the  whole body  of  information (the  whole  ``M'') suddenly
appears.

This intuitive similarity of collapse events to conscious events is a
reason why many quantum theorists are attracted to the idea that conscious
events are quantum events. Orthodox quantum theory rests on that idea.

There is in the  quantum ontology a  tie-in to past and  future, because  if
one asks what the present reality is,  the answer can be either knowledge of
the  past, or   potentialities for  the  future: the  present  is  an abrupt
transition from fixed past to open future, not a slice of a  self-sufficient
continuous reality.

\newpage
\noindent {\bf Appendix D. The Dilemma of Free Will.}
\vskip 0.1in

The two horns of this dilemma are `determinism' and `chance'. If determinism
holds  then  a  person  seems   reduced to  a   mechanical  device,  no more
responsible for his acts  than a clock is  responsible for telling the wrong
time. But if  determinism fails then  his actions are  controlled in part by
``chance'', rendering him {\it even less} responsible for his acts.

This   argument  can   powerfully  affect  on our   lives: it   allows us to
rationalize our  own moral failings,  and it influences  the way we, and our
institutions, deal with the failings of others.

It might appear that there is no way  out: either the world is deterministic
or it's not, and the  second possibility involves  chance. So we get hung on
one horn or the other. 

Quantum ontology evades both horns.

The point is that determinism does not imply mechanism. The reason we say we
are not responsible if determinism  holds is that ``determinism'' evokes the
idea of  ``mechanism''; it evokes the  idea of a clock.  And, indeed, that's
exactly what {\it  is} entailed by  the determinism of  classical mechanics.
According to the principles of classical mechanics everything you will do in
your life was fixed and settled  before you were born {\it by local `myopic'
mechanical  laws}: i.e., by  essentially the  same sort of  local mechanical
linkages that  control the workings  of a clock. If your  thoughts and ideas
enter  causally into  the  physical  proceedings at  all, it is  only to the
extent  that  they are   themselves  completely   controlled by  these local
mechanical processes.  Hence the causes of your  actions can be reduced to a
huge assembly of thoughtless microscopic processes.

But in quantum dynamics  our knowings enter as  the central dynamical units.
What we have is a dynamics of knowings that evolve according to the rules of
quantum dynamics.  To be sure these  dynamical rules do  involve elements of
chance, but these are no more problematic than the thermal and environmental
noise that occurred in the classical  case: our high-level structures cannot
maintain total  fine control  over every  detail. But there  is, in spite of
that important  similarity, a huge difference  because in the classical case
everything   was   determined  from the  bottom  up,  by   thoughtless micro
processes, whereas in the quantum case everything is determined from the top
down, by a dynamics that connects earlier knowings to later knowings.

And these knowings are doing what we feel they are doing: initiating complex
actions, both physical and mental, that pave the way to future knowings.

No reduction  to knowingless  process is  possible because  each step in the
dynamical  processes is the  actualization of a knowing  that is represented
mathematically as the grasping, as a  whole, of a structural complex that is
equivalent to the structure of the knowing.

\end{document}